\documentclass[prd,twocolumn,eqsecnum,showpacs,amsfonts,amssymb]{revtex4}

\usepackage{graphicx}

\usepackage{bm}

\newcommand{\beq}{\begin{equation}}
\newcommand{\eeq}{\end{equation}}
\newcommand{\beqs}{\begin{eqnarray}}
\newcommand{\eeqs}{\end{eqnarray}}

\newcommand{\drawsquare}[2]{\hbox{%
\rule{#2pt}{#1pt}\hskip-#2pt%  left vertical
\rule{#1pt}{#2pt}\hskip-#1pt%  lower horizontal
\rule[#1pt]{#1pt}{#2pt}}\rule[#1pt]{#2pt}{#2pt}\hskip-#2pt%  upper horizontal
\rule{#2pt}{#1pt}}% right vertical
\newcommand{\fund}{\raisebox{-.5pt}{\drawsquare{6.5}{0.4}}}%  fund
\newcommand{\sym}{\raisebox{-.5pt}{\drawsquare{6.5}{0.4}}\hskip-0.4pt%
        \raisebox{-.5pt}{\drawsquare{6.5}{0.4}}}%  symmetric second rank
\newcommand{\asym}{\raisebox{-3.5pt}{\drawsquare{6.5}{0.4}}\hskip-6.9pt%
        \raisebox{3pt}{\drawsquare{6.5}{0.4}}}%  antisymmetric second rank

\begin{document}

\title{Higher-Loop Calculations of the 
Ultraviolet to Infrared Evolution of a Vectorial Gauge 
Theory in the Limit $N_c \to \infty$, $N_f \to \infty$ with 
$N_f/N_c$ Fixed}

\author{Robert Shrock}

\affiliation{C. N. Yang Institute for Theoretical Physics,
Stony Brook University, Stony Brook, NY 11794}

\begin{abstract}

We consider an asymptotically free vectorial SU($N_c$) gauge theory with $N_f$
fermions in the fundamental representation and analyze higher-loop
contributions to the evolution of the theory from the ultraviolet to the
infrared in the limit where $N_c \to \infty$ and $N_f \to \infty$ with
$r=N_f/N_c$ a fixed, finite constant.  We focus on the case where the
$n$-loop beta function has an infrared zero, at $\xi=\xi_{IR,n\ell}$, where
$\xi=\alpha N_c$. We give results on $\xi_{IR,n\ell}$, the anomalous
dimension of the fermion bilinear evaluated at $\xi_{IR,n\ell}$, denoted
$\gamma_{IR,n\ell}$, and certain structural properties of the beta function,
$\beta_\xi$.  The approach to this limit is investigated, and it is shown that
the leading correction terms are strongly suppressed, by the factor $1/N_c^2$.
This provides an understanding of a type of approximate universality in
calculations for moderate values of $N_c$ and $N_f$, namely that
$\alpha_{IR,n\ell}N_c$, $\gamma_{IR,n\ell}$, and structural properties of the
beta function are similar in theories with different values of $N_c$ and $N_f$
provided that they have similar values of $N_f/N_c$.  We give results up to
four loops for nonsupersymmetric theories and up to three loops for 
supersymmetric theories.

\end{abstract}

\pacs{11.15.-q,11.10.Hi,11.15.Pg}

\maketitle

% =======================================================================

\section{Introduction}
\label{intro}

The evolution of an asymptotically free gauge theory from the 
ultraviolet (UV) to the infrared (IR) is of fundamental field-theoretic 
interest. The UV to IR evolution of the gauge coupling $g(\mu)$ as a function
of the Euclidean momentum scale, $\mu$, is determined by the $\beta$ function
\cite{beta} 
\beq
\beta_\alpha \equiv \frac{d\alpha}{dt} \ , 
\label{betadef}
\eeq
where $t=\ln \mu$ and $\alpha(\mu)=g(\mu)^2/(4\pi)$.  Here we consider this
evolution for a vectorial gauge theory with gauge group $G={\rm SU}(N_c)$ and
$N_f$ massless fermions $\psi_j$, $j=1,...,N_f$, transforming according to the
fundamental representation of $G$ \cite{fm}.  We also point out some contrasts
with results for fermions in higher-dimensional representations.  We focus on
the case where the $n$-loop $\beta$ function has an infrared zero, at a value
$\alpha = \alpha_{IR,n\ell}$.  The condition of asymptotic freedom requires
that $N_f$ be bounded above by a value $N_{f,b1z}$ where the one-loop
coefficient in $\beta$ vanishes \cite{b1}.  For large enough $N_f$ (less than
$N_{f,b1z}$), the two-loop $\beta$ function has an infrared zero at a certain
value of $\alpha$, denoted $\alpha_{IR,2\ell}$ \cite{b2,bz}.  The desire to
understand better both the behavior of the running coupling in quantum
chromodynamics (QCD) and the properties of an IR zero that occurs for
sufficiently large $N_f$ have motivated calculations of higher-loop terms in
$\beta$ \cite{b3,b4} and higher-loop corrections to the two-loop result for the
IR zero \cite{gkgg}-\cite{bc}. In \cite{bvh,ps}, calculations of the IR zero in
$\beta$, and the associated anomalous dimension of the (gauge-invariant)
fermion bilinear, $\gamma_m$, were done to four-loop order for an
asymptotically free vectorial gauge theory with gauge group $G$ and $N_f$
fermions in an arbitrary representation $R$, with explicit results for $R$
equal to the fundamental, adjoint, and symmetric and antisymmetric rank-2
tensor representations.  Further generalizations and results on higher-loop
structural properties of the $\beta$ function were given in \cite{bc}.

An interesting and important property that one notices in these calculations
for an SU($N_c$) theory with $N_f$ massless fermions in the fundamental
representation is that the values of the $n$-loop $\gamma_m$, evaluated at
$\alpha_{IR,n\ell}$, denoted $\gamma_{IR,n\ell}$, and of the product
$\alpha_{IR,n\ell}N_c$, are similar for theories with different values of $N_c$
and $N_f$, provided that these theories have similar values of the ratio
$N_f/N_c$. Indeed, the computations in \cite{bc} show that this is also true
for other structural quantities describing the UV to IR evolution, including
the derivative of the $n$-loop beta function, $d\beta_{n\ell}/d\alpha$
evaluated at $\alpha_{IR,n\ell}$ and the products $\alpha_{m,n\ell}N_c$ and
$(\beta_{n\ell})_{min}N_c$, where $\alpha_{m,n\ell}$ denotes the value of
$\alpha$ where $\beta_{n\ell}$ is a minimum, and $(\beta_{n\ell})_{min}$ is the
value of $\beta_{n\ell}$ at this minimum.  These observations show that there
is an underlying approximate universality in the form of the quantities that
control the UV to IR evolution of these theories.  This motivates a more
detailed study to elucidate this phenomenon. We carry out this study in the
present work.

For this purpose, we analyze these theories in the 't Hooft - Veneziano limit
\cite{thooftlargeN,veneziano}
\beqs
& & N_c \to \infty \ , \quad N_f \to \infty \cr\cr
& & {\rm with} \ r \equiv \frac{N_f}{N_c} = \kappa \quad 
{\rm and} \quad \xi(\mu) \equiv \alpha(\mu) N_c = \lambda(\mu) \ , \cr\cr
& & 
\label{lnn}
\eeqs
where $\kappa$ is a constant and $\lambda$ is a function depending only on
$\mu$. We will use the symbol $\lim_{LNN}$ for this limit, where ``LNN'' stands
for ``large $N_c$ and $N_f$'' (with the constraints in Eq. (\ref{lnn})
imposed).  The reasons for these two constraints in Eq. (\ref{lnn}) (that the
ratio $N_f/N_c$ and the product $\alpha(\mu)N_c$ are fixed and finite) are that
these constitute the necessary and sufficient conditions in order that (i)
fermions give nonvanishing contributions to the $\beta$ function, anomalous
dimension $\gamma_m$, and other quantities, and (ii) scattering amplitudes
remain finite, in the limit $N_c \to \infty$, respectively.  More generally, if
the fermions are nonsinglets under other gauge groups with squared couplings
$\alpha_i$, one also requires that the products $\alpha_i N_c$ be finite as 
$N_c \to \infty$ \cite{lnc}.

As we will show in detail below, a study of the LNN limit (\ref{lnn}) and the
approach to it provides an explanation of the approximate universality in the
UV to IR evolution of theories with different values of $N_c$ and $N_f$ but 
the same, or similar, values of $r$ that is exhibited in explicit
calculations (with appropriate scalings understood for certain quantities, such
as multiplying $\alpha_{IR,n\ell}$ by $N_c$).  A crucial property of the LNN
limit is that it reduces the number of variables on which the UV to IR
evolution depends.  Thus, for finite $N_c$ and $N_f$, this evolution and the
$\beta$ function that describes it, depend on three variables: $\alpha$ (and
thus, parametrically, $\mu$), $N_c$, and $N_f$, while in the LNN limit, they
only depend on the two variables $\alpha(\mu)$ and $r$.

Since the rational numbers ${\mathbb Q}$ are dense in the real numbers
${\mathbb R}$, it follows that in the LNN limit, one can choose values of $N_c$
and $N_f$ so that the rational number $r$ is arbitrarily close to any
non-negative real number.  Therefore, henceforth, to arbitrarily good accuracy,
we may simply treat $r$ as a real number, and we will do so. As is well-known,
in the $N_c \to \infty$ limit and also in the LNN limit, the gauge group
SU($N_c)$ is effectively equivalent to U($N_c$).  The use of a large-$N$ limit,
where $N$ is the number of components in a spin or field, has been valuable in
the past partly because it allowed one to obtain exact results for statistical
mechanical models \cite{stanley} and quantum field theories
\cite{thooftlargeN}-\cite{lnc}, \cite{earlylargeN,largeNreview}.  Our purpose
in using it here is somewhat different, namely to gain further insight into the
above-mentioned approximate universality that is exhibited by calculations of
the UV to IR evolution of theories with different $N_c$ and $N_f$ but equal or
similar values of $r$.

  As part of our analysis, we will briefly contrast the properties of theories
with fermions in the fundamental representation with properties of theories
with fermions in higher-dimensional representations.  In the case where
fermions are in a two-index representation (including the adjoint, and
symmetric and antisymmetric rank-2 tensor representations), the condition that
is necessary and sufficient to construct a finite $N_c \to \infty$ limit is to
set $N_f$ equal to a (non-negative, integer) constant. This is a result of the
fact that the quadratic Casimir invariants $T_f$ and $C_f$ \cite{casimir} that
enter into the coefficients of the beta function grow like (a constant times)
$N_c$ as $N_c \to \infty$. If the fermions are in a representation involving
three or more indices, then generically for fixed finite $N_f$, the fermion
contributions dominate over the gauge field contributions to the $\beta$
function by powers of $N_c$ as $N_c \to \infty$.  For example, for the
symmetric rank-3 tensor representation, $T_f \sim N_c^2$ for large $N_c$, so
that the fermion contribution to the leading $\beta$ function coefficient
dominates over the gauge-field contribution, which is $\sim N_c$, spoiling the
asymptotic freedom of the theory.  Hence, aside from our primary focus on the
case of fermions in the fundamental representation, we will restrict our
discussion of other representations to the adjoint, and symmetric and
antisymmetric rank-2 tensors.

By taking $r$ near to its maximum value allowed by asymptotic freedom, one can
arrange that the zero of $\beta$ occurs at an arbitrarily small value of
$\xi_{IR}$, and one may conclude that in the infrared the theory is in a
deconfined non-Abelian Coulomb phase without any spontaneous chiral symmetry
breaking.  In this case, the IR zero of $\beta$ at $\xi_{IR}$ is an exact fixed
point of the renormalization group for the theory.  In contrast, as $r$
decreases, $\xi_{IR}$ increases, and studies with finite $N_c$ and $N_f$ lead
to the conclusion that for $N_f$ less than a critical value, $N_{f,cr}$, as
$\mu$ decreases though a value denoted $\Lambda$, the interaction strength
$\alpha(\mu)$ exceeds a critical value, $\alpha_{cr}$, to produce spontaneous
chiral symmetry breaking and associated dynamical mass generation for the
fermions.  For a given $N_c$, the theory may thus be considered to undergo a
(zero-temperature) chiral phase transition as $N_f$ passes through this value,
$N_{f,cr}$ \cite{ap}, and there has been an intensive research program using
lattice gauge simulations to determine $N_{f,cr}$ for values such as $N_c=3$
and $N_c=2$ \cite{conf}.  Correspondingly, in the LNN limit considered here,
the theory undergoes a chiral phase transition as $r$ passes through $r_{cr}$,
where $r_{cr} = N_{f,cr}/N_c$, with the UV to IR evolution leading to a
chirally symmetric phase for $r > r_{cr}$ and a phase with spontaneous chiral
symmetry breaking for $r < r_{cr}$.  If $r < r_{cr}$, then in the effective
low-energy field theory below $\Lambda$, one integrates out the fermions (which
have dynamically generated masses of order $\Lambda$), and the $\beta$ function
changes to become that of a pure non-Abelian gauge theory, which does not have
a (perturbative) zero.  In this case, $\xi_{IR}$ is only an approximate fixed
point.

This paper is organized as follows.  In Section \ref{betafunction} we define an
appropriately scaled beta function, called $\beta_\xi$, that is a finite
function of $\xi$ in the LNN limit and in this section, and in Sections
\ref{irzero} and \ref{structural} we investigate its structure up to four-loop
order.  Our results include an analysis of the behavior of the coefficients as
functions of $r$, the LNN limits for the $n$-loop IR zero, $\xi_{IR,n\ell}$,
the value of $\xi$ where $\beta_\xi$ is a minimum, the value of $\beta_\xi$ at
this minimum, and the derivative $d\beta_\xi/d\xi$ evaluated at
$\xi_{IR,n\ell}$.  In Section \ref{anomdim} we carry out a similar analysis of
the coefficients in the anomalous dimension of the fermion bilinear,
$\gamma_m$, and its value at $\xi_{IR,n\ell}$, again up to
four-loop order.  In Section \ref{approach} we calculate correction terms to
the LNN limits for various quantities and give a general analytic explanation
for the rapidity with which this limit is approached, namely that these
correction terms are strongly suppressed, by the factor $1/N_c^2$.  Section
\ref{susy} is devoted to a corresponding study of the LNN limit of a
supersymmetric gauge theory.  Our conclusions are contained in a final section.

% ======================================================================

\section{$\beta$ Function and Some General Properties in the LNN Limit} 
\label{betafunction}

\subsection{General} 

In this section we analyze the $\beta$ function in the limit (\ref{lnn}).  It
will be convenient to 
define $a(\mu) \equiv g(\mu)^2/(16\pi^2) = \alpha(\mu)/(4\pi)$ and 
\beq
x(\mu) = \frac{\xi (\mu)}{4\pi} \ . 
\label{x}
\eeq
(The argument $\mu$ will often be suppressed in the notation.)  In terms of 
$\alpha$, or equivalently, $a$, the beta function has the series expansion
\beq
\beta \equiv \beta_\alpha = -8\pi a \sum_{\ell=1}^\infty b_\ell \, a^\ell = 
 -2\alpha \sum_{\ell=1}^\infty \bar b_\ell \, \alpha^\ell \ , 
\label{beta}
\eeq
where $\ell$ denotes the loop order and $\bar b_\ell \equiv
b_\ell/(4\pi)^\ell$. Thus, the $n$-loop beta function is given by
Eq. (\ref{beta}) with $\infty$ replaced by $n$ as the upper limit on the
summation over loop order, $\ell$.  The coefficients $b_\ell$ for $\ell=1, \ 2$
are independent of the scheme used for the regularization and renormalization
of the theory and were calculated in \cite{b1} and \cite{b2}.  The $b_\ell$
with $\ell \ge 3$ are scheme-dependent and have been calculated up to
$(\ell=4)$-loop order \cite{b3,b4} in the modified minimal subtraction
\cite{ms} ($\overline{MS}$) scheme \cite{msbar}.  The usefulness of the
$\overline{MS}$ scheme has been demonstrated, e.g., by the fact that inclusion
of three-loop and four-loop corrections in the running of $\alpha_s(\mu)$ in
QCD significantly improves the fit to experimental data \cite{bethke}.  The
scheme-dependence of the higher-loop IR zero of the beta function was recently
studied in \cite{sch}.

For our present analysis of the theory in the LNN limit, the first step is to
construct a beta function that has a finite, nontrivial LNN limit.  We do this
by multiplying both sides of (\ref{beta}) by $N_c$ and then taking the LNN
limit.  The result is a function of $\xi$ and can be expressed as 
\beq
\beta_{\xi} \equiv \frac{d\xi}{dt} = \lim_{LNN} \beta_\alpha N_c \ . 
\label{betaxi}
\eeq
This function has the expansion 
\beq
\beta_\xi \equiv \frac{d\xi}{dt} 
= -8\pi x \sum_{\ell=1}^\infty \hat b_\ell x^\ell
= -2 \xi \sum_{\ell=1}^\infty \tilde b_\ell \xi^\ell \ , 
\label{betaxiseries}
\eeq
where
\beq
  \hat b_\ell = \lim_{LNN} \frac{b_\ell}{N_c^\ell} \ , \quad 
\tilde b_\ell = \lim_{LNN} \frac{\bar b_\ell}{N_c^\ell} \ .
\label{bellrel}
\eeq
Thus, similarly to the relation between $\bar b_\ell$ and $b_\ell$,
\beq
\tilde b_\ell = \frac{\hat b_\ell}{(4\pi)^\ell} \ . 
\label{tildehatbell}
\eeq
As with Eq. (\ref{bellrel}), it is understood here and below that all
expressions have been evaluated in the LNN limit (\ref{lnn}).  The $\beta_\xi$
function, calculated to $n$-loop ($n\ell$) order, is denoted by
$\beta_{\xi,n\ell}$ and is given by Eq. (\ref{betaxiseries}) with $\infty$
replaced by $n$ as the upper limit on the sum over $\ell$.

To analyze the zeros of $\beta_{\xi,n\ell}$, aside
from the double zero at $\xi=x=0$, we extract an overall factor of
$-2\xi^2$ and calculate the zeros of the reduced $(r)$ polynomial
\beq
\beta_{\xi,n\ell,r} \equiv
-\frac{\beta_{\xi,n\ell}}{2 \xi^2} = 
\sum_{\ell=1}^n \tilde b_\ell \, \xi^{\ell-1} = 
\frac{1}{4\pi}\sum_{\ell=1}^n \hat b_\ell \, x^{\ell-1} \ . 
\label{betaxi_nloop_reduced}
\eeq
As is clear from Eq. (\ref{betaxi_nloop_reduced}), the zeros of
$\beta_{\xi,n\ell}$ away from the origin depend only on $n-1$ ratios of
coefficients, which can be taken as $\tilde b_\ell/\tilde b_n$ for
$\ell=1,...,n-1$.  Although Eq. (\ref{betaxi_nloop_reduced}) is an algebraic
equation of degree $n-1$, with $n-1$ roots, only one of these is physically
relevant as the IR zero of $\beta_{\xi,n\ell}$. We denote this as
$\xi_{IR,n\ell}=4 \pi x_{_{IR,n\ell}}$.

% ===================================================================

\bigskip

\subsection{Behavior of Coefficients in $\beta$ as Functions of $r$}

From the expressions for $b_1$ and $b_2$ \cite{b1,b2}, we have
\beq
\hat b_1 = \frac{1}{3}(11-2r)
\label{b1hat}
\eeq
and
\beq
\hat b_2 = \frac{1}{3}(34-13r)  \ . 
\label{b2hat}
\eeq
In the $\overline{MS}$ scheme, from the expression for $b_3$ \cite{b3}, we
obtain 
\beqs
\hat b_3 & = & \frac{1}{54}(2857-1709r+112r^2) \cr\cr
& = & 52.9074-31.6481r+2.07407r^2
\label{b3hat}
\eeqs
and from $b_4$ \cite{b4}, we obtain 
\begin{widetext}
\beqs
\hat b_4 & = & \frac{150473}{486}- 
 \Big ( \frac{485513}{1944} \Big ) r
+\Big ( \frac{8654}{243} \Big ) r^2 
+\Big ( \frac{130}{243}  \Big ) r^3 + \frac{4}{9}(11-5r+21r^2) \zeta(3) 
\cr\cr
& = & 315.492 - 252.421 \, r + 46.832 \, r^2 + 0.534979 \, r^3 \ , 
\label{b4hat}
\eeqs
\end{widetext}
to the indicated numerical floating-point accuracy, where $\zeta(s) =
\sum_{n=1}^\infty n^{-s}$ is the Riemann $\zeta$ function.
For some purposes it is more convenient to deal with the $\hat
b_\ell$, since they are free of factors of $4 \pi$, while for numerical
purposes it is often more convenient to use the $\tilde b_\ell$, since the
range of values of $\tilde b_\ell$ as functions of $\ell$ is somewhat smaller
than the range for the $\hat b_\ell$.  In Table \ref{btilde_nloop_values} we
list values of $\tilde b_\ell$ for $2 \le \ell \le 4$ as functions of $r$ in
the interval $0 \le r \le r_{b1z}$.

In \cite{bvh}, $N_{f,b \ell z}$ was defined as the value or set of values 
of $N_f$ where $b_\ell=0$ and thus, for our present analysis, we define 
\beq
r_{b \ell z} \equiv \lim_{LNN} \frac{N_{f,b \ell z}}{N_c} \ . 
\label{rbjz}
\eeq
From Eq. (\ref{b1hat}), we have 
\beq
r_{b1z} = \frac{11}{2} \ . 
\label{rb1zhvl}
\eeq

As is evident from Eqs. (\ref{b1hat}) and (\ref{b2hat}), the coefficients $\hat
b_1$ and $\hat b_2$ are both monotonically (and linearly) decreasing functions
of $r$. The coefficient $\hat b_1$ decreases from 11/3 to 0 as $r$ increases
from 0 to $r_{b1z} = 11/2$.  We require that the theory be asymptotically free,
i.e., 
\beq
r < r_{b1z} = \frac{11}{2} \ .  
\label{rrange}
\eeq

The coefficient $\hat b_2$ decreases from 34/3 at $r=0$ and passes through zero
to negative values as $r$ increases through the value 
\beq
r_{b2z} = \frac{34}{13} \ . 
\label{rb2zhvl}
\eeq
As $r \nearrow r_{b1z}$, $\hat b_2$ reaches the value 
\beq
\hat b_2 = - \frac{25}{2} \quad {\rm at} \ \ r = r_{b1z} \ . 
\label{b2hatrb1z}
\eeq
Therefore, in the LNN limit, the interval in $r$, denoted $I_r$, where the 
two-loop $\beta_\xi$ function has an IR zero, is given by
\beq
I_r: \quad r_{b2z} < r < r_{b1z} \ , \quad 
\frac{34}{13} < r < \frac{11}{2} 
\label{rinterval}
\eeq
(i.e., $2.615 < r < 5.500$).  
With $r \in I_r$, we will, correspondingly, focus on the
UV to IR evolution in the interval
\beq
I_\xi: \quad 0 \le \xi(\mu) \le \xi_{IR,n\ell} \ . 
\label{xiinterval}
\eeq

The coefficient $\hat b_3$ vanishes at two values of $r$, denoted 
\beq
r_{b3z1} = \frac{1709 - 57\sqrt{505}}{224} = 1.911
\label{rb3z1}
\eeq
and
\beq
r_{b3z2} = \frac{1709 + 57\sqrt{505}}{224} =  13.348 \ , 
\label{rb3z2}
\eeq
where here and below, the floating-point values are given to the indicated
accuracy.  This coefficient $\hat b_3$ is monotonically decreasing in the
interval $0 < r < r_{b1z}$, decreasing from $\hat b_3=2857/54 = 52.907$ at
$r=0$ and passing through zero to negative values as $r$ increases through
$r_{b3z1}$ in Eq. (\ref{rb3z1}). As $r$ increases from $r_{b3z1}$, $\hat b_3$
continues to decrease and passes through the value
\beq
\hat b_3 = -\frac{5299}{338} = -15.6775 \quad {\rm at} \ \ r= r_{b2z} = 
\frac{34}{13} 
\label{b3nfb2z}
\eeq
at the lower end of the interval $I_r$. As $r$ increases throughout the
interval $I_r$, $\hat b_3$ decreases further, and as $r$ increases to its
maximum, $r_{b1z}$, at the upper end of this interval, $\hat b_2$ reaches 
the value 
\beq
\hat b_3 = -\frac{701}{12} = -58.417 \quad{\rm at} \ \ r = r_{b1z} = 
\frac{11}{2} \ . 
\label{b3hvrb1z}
\eeq
Since
\beq
r_{b3z1} < r_{b2z}
\label{rb3zminus_lt_rb2z}
\eeq
and
\beq
r_{b3z2} > r_{b1z} \ , 
\label{rb3zplus_gt_rb1z}
\eeq
it follows that in the $\overline{MS}$ scheme, 
\beq
\hat b_3 < 0 \quad \forall \ \ r \in I_r \ .
\label{b3negative_in_interval_r}
\eeq
Given this result and the fact that the quantity $-54\hat b_3$ will appear in
later formulas, it will be convenient to denote 
\beq
D_{3\ell} \equiv -54 \hat b_3 = -2857+1709r-112r^2 \ , 
\label{d3ell}
\eeq
which is positive for $r \in I_r$.

For completeness, we note that $\hat b_3$ reaches a minimum at $r=1709/224 =
7.629$, and, for larger $r$, it increases, passing through zero again at the
value $r_{b3z2}$ in Eq. (\ref{rb3z2}).  Since these values of $r$ lie above
$r_{b1z}$, they are not of direct interest for our present study.

As $r$ increases through the range $0 \le r < r_{b1z}$, the coefficient $\hat
b_4$ in the $\overline{MS}$ scheme decreases from the value
\beq
\hat b_4 = \frac{150473}{486} + \frac{44\zeta(3)}{9} = 315.492 \quad 
{\rm at} \ \ r=0 \ , 
\label{b4hatr0}
\eeq
(i.e., $\tilde b_4 = 1.265 \times 10^{-2}$), passes through zero with negative
slope at
\beq
r_{b4z,1}=2.040 \ ,
\label{rb4z1}
\eeq
reaches a minimum of  
$-14.831$ at $r=2.581$, and then increases.  At the lower end of the interval
$I_r$, at $r_{b2z}$, 
\beq
\hat b_4 = -\frac{550009}{6084} + \frac{31900 \, \zeta(3)}{507} = 
-14.770 \quad {\rm at} \ \ r=r_{b2z} \ . 
\label{b4hat_at_rb2z}
\eeq
As $r$ increases in the interval $I_r$, $\hat b_4$ passes through zero again, 
at
\beq
r_{b4z,2}=3.119 \ , 
\label{rb4z2}
\eeq
this time with positive slope, and attains the value 
\beq
\hat b_4 = \frac{14731}{144}+275\zeta(3) = 432.864 \quad {\rm at} \ \ 
r=r_{b1z} 
\label{b4hatrb1z}
\eeq
(i.e., $\tilde b_4=1.736 \times 10^{-2}$) at the upper end of the interval
$I_r$.  Some special values, expressed in terms of the $\tilde
b_\ell$ coefficients, are listed in Table \ref{btilde_nloop_specialvalues}. 

Concerning the sign of $\hat b_4$ in the range $r \ge 0$, 
\beqs 
& & \hat b_4 > 0 \quad {\rm for} \ 0 < r < r_{b4z,1} \ , \cr\cr
& & \hat b_4 < 0 \quad {\rm for} \ r_{b4z,1} < r < r_{b4z,2} \ ,  \cr\cr
& & \hat b_4 > 0 \quad {\rm for} \ r_{b4z,1} < r < r_{b1z} \ . 
\label{b4sign}
\eeqs
That is, numerically, $\hat b_4 > 0$ \ if \ $0 < r < 2.040$ or 
$r > 3.119$, and $b_4 < 0$ if $2.040 < r < 3.119$.
The zero of $\hat b_4$ at $r_{b4z,1}=2.040$ lies
below the lower end of the interval $I_r$ (at $r_{b2z}=2.615$), while the zero
at $r_{b4z,2}=3.119$ lies in the interior of the interval $I_r$. 
Hence, restricting to $r \in I_r$, 
\beqs
{\rm For} \ r \in I_r \ , \quad & & \hat b_4 < 0 \ \ {\rm if} \ \
r_{b2z} < r < r_{b4z,1} \cr\cr
& & \hat b_4 > 0  \ \ {\rm if} \ \ r_{b4z,1} < r < r_{b1z} \ , 
\cr\cr
& &
\label{b4sign_in_rinterval}
\eeqs
i.e., numerically, for $r \in I_r$, $\hat b_4 < 0$ if $2.615 < r < 3.119$ 
and $\hat b_4 > 0$ if $3.119 < r < 5.500$. 
Since $\hat b_4$ is a cubic polynomial in $r$, there is a third value of $r$
where it vanishes, but this is at the negative, and hence unphysical, value
$r=-92.699$ and hence is of no direct relevance here.  Although our analysis
here presumes the LNN limit, a remark is in order for finite $N_c$ and $N_f$.
The interval where $\hat b_4$ is negative for $r \in I_r$ is not present for
sufficiently small $N_c$ and $N_f$.  This is evident from the explicit $\bar
b_4$ values listed in Table I of our Ref. \cite{bvh} for $N_c=2$ and
$N_c=4$. This interval of negative $\bar b_4$ values is present for $N_c \ge
4$.

% =======================================================================

\section{IR Zero of $\beta$}
\label{irzero}

Combining the results from the previous section, we exhibit the explicit 
four-loop $\beta_\xi$ function. For this
purpose, it is simplest to use the $x$ variable defined in Eq. (\ref{x}). We
have 
\begin{widetext}
\beqs
& & \beta_\xi = -8\pi x^2 \bigg [ \ \frac{11-2r}{3} 
+ \Big ( \frac{34-13r}{3} \Big )x 
+ \Big ( \frac{2857-1709r+112r^2}{54} \Big )x^2 \cr\cr
& & 
+ \bigg \{ \frac{150473}{486}-  \Big ( \frac{485513}{1944} \Big ) r
+\Big ( \frac{8654}{243} \Big ) r^2
+\Big ( \frac{130}{243}  \Big ) r^3 + \frac{4}{9}(11-5r+21r^2) \zeta(3) \bigg
\} x^3 + O(x^4) \ \bigg ] \ . 
\cr\cr
& & 
\label{betaxi_4loop}
\eeqs
\end{widetext}
%

% ===================================================================

\subsection{Two-Loop Level}

At the two-loop level, if $r \in I_r$, then $\beta_\xi$ has an IR zero at
\beqs
\xi_{IR,2\ell} & = & -\frac{\tilde b_1}{\tilde b_2} = 
-\frac{4\pi \hat b_1}{\hat b_2} \cr\cr
& = & \frac{4\pi(11-2r)}{13r-34} \ .
\label{xiir_2loop}
\eeqs
Since this is obtained from a perturbative calculation, it is only reliable if
$\xi_{IR,2\ell}$ is not too large.  As $r \to r_{b1z}$, $\xi_{IR,2\ell}
\to 0$, and hence in the upper end of the interval $I_r$, one may plausibly
expect that this two-loop expression becomes a progressively more and more
accurate approximation to the IR zero of the exact $\beta_\xi$ function.
As $r \searrow 34/13$ at the lower end of the interval $I_r$,
$\xi_{IR,2\ell}$ grows too large for this perturbative calculation to be
applicable.  It will be useful here and below to give values of various
quantities at an illustrative value of $r$. 

% ===================================================================

\subsection{Three-Loop Level} 

At the three-loop level, the IR zero of $\beta_\xi$ is given by the physical 
(smallest positive) root of the quadratic equation
\beq
\beta_{\xi,2\ell,r} = \tilde b_1 + \tilde b_2 \xi + \tilde b_3 \xi^2 = 0 \ . 
\label{eqxi_3loop}
\eeq
This equation has, formally, two solutions, namely
\beq
\frac{1}{2\tilde b_3}\Big ( -\tilde b_2 \pm \sqrt{\tilde b_2^2 -
4 \tilde b_1 \tilde b_3} \ \Big ) \ .
\label{xiir_3loopboth}
\eeq
Since we have shown that $\tilde b_3 < 0$ for $r \in I_r$ for a general scheme
that preserves the existence of the IR zero in the (scheme-independent) 
$\beta_{\xi,2\ell}$ at the three-loop level, we can rewrite 
(\ref{xiir_3loopboth}) as 
\beq
\frac{1}{2 |\tilde b_3|}\Big ( -|\tilde b_2| \mp 
\sqrt{\tilde b_2^2 + 4 \tilde b_1 |\tilde b_3|} \ \Big ) \ . 
\label{xiir_3loopboth2}
\eeq
As is evident from Eq. (\ref{xiir_3loopboth2}), only the root corresponding to
the lower sign choice in Eq. (\ref{xiir_3loopboth2}) is
positive and hence physical.  We denote it as 
\beq
\xi_{IR,3\ell} = \frac{1}{2 |\tilde b_3|}\Big ( -|\tilde b_2| + 
\sqrt{\tilde b_2^2 + 4 \tilde b_1 |\tilde b_3|} \ \Big ) \ . 
\label{xiir_3loop}
\eeq

By the same type of proof as was given in \cite{bvh,bc}, for the relevant
interval $r \in I_r$ where the scheme-independent two-loop $\beta_{\xi,2\ell}$
function has an IR zero, we find that at the three-loop level
\beq
\xi_{IR,3\ell} \le \xi_{IR,2\ell} \quad {\rm for} \ \ r \in I_r \ , 
\label{xiir_2versus3loop}
\eeq
with equality only at $r=r_{b1z}$, where $\xi_{IR,3\ell}=\xi_{IR,2\ell}=0$.  In
\cite{bc} we pointed out that the corresponding inequality $\alpha_{IR,3\ell} <
\alpha_{IR,2\ell}$ applies more generally than just in the $\overline{MS}$
scheme, and the same is true of the inequality (\ref{xiir_2versus3loop}).  We
recall the reasoning for this. Since the existence of an IR zero in the
two-loop $\beta$ function, $\beta_{\xi,2\ell}$, is a scheme-independent
property of the theory, a reasonable scheme should maintain the existence of
this IR zero (albeit with a shifted value) at higher-loop order.  Now in order
for a scheme to maintain this zero, a necessary and sufficient condition is
that $\tilde b_2^2 - 4 \tilde b_1 \tilde b_3 \ge 0$, so that the square root in
Eq. (\ref{xiir_3loopboth}) is real.  But the lower end of the interval $I_r$ is
defined by the condition that $\tilde b_2 \to 0$ as $r \searrow r_{b2z}$. Given
that $r \in I_r$ so $\beta_{\xi,2\ell}$ has an IR zero, this means that a
reasonable scheme, which preserves the existence of this zero at the three-loop
level, should have $\tilde b_3 < 0$ for $r \in I_r$.  From this, by the same
type of proof as was given in \cite{bc} for this class of schemes, the
inequality (\ref{xiir_2versus3loop}) follows.

Substituting the relevant expressions for the $\tilde b_\ell$ in
(\ref{xiir_3loop}), we have, in the $\overline{MS}$ scheme, the explicit 
result
\beq
\xi_{IR,3\ell}=\frac{12\pi [-3(13r-34) + \sqrt{C_{3\ell}} \ ]}{D_{3\ell}} \ , 
\label{xiir_3loop_explicit}
\eeq
where $D_{3\ell}$ was defined above in Eq. (\ref{d3ell}), and it is convenient 
to define the shorthand notation
\beq
C_{3\ell} = -52450+41070r-7779r^2+448r^3 \ . 
\label{c3ell}
\eeq
The polynomial $C_{3\ell}$ has only one real zero, at $r=1.86532$ (to the
indicated accuracy) and is positive for $r > 1.86532$, and hence
for all $r \in I_r$.  The polynomial $-3(13r-34)$ vanishes at the lower end of
the interval $I_r$ and is negative for $r \in I_r$, but it is smaller than
$\sqrt{C_{3\ell}}$, so $\xi_{IR,3\ell} > 0$ for $r \in I_r$, as is necessary
for it to be physical. 

In Table \ref{xiir_nloop_values} we list numerical values of $\xi_{IR,2\ell}$
and $\xi_{IR,3\ell}$ for $r \in I_r$. As is evident in this table,
$\xi_{IR,2\ell}$ and $\xi_{IR,3\ell}$ decrease monotonically as a function of
$r$ throughout this interval $I_r$ (as does $\xi_{IR,4\ell}$, to be discussed
below). At the lower end of this interval,
\beq
\xi_{IR,3\ell} = 20\pi \sqrt{\frac{26}{5299}} = 4.401 \quad {\rm at} \
r=\frac{34}{13} \ ,
\label{xiir_3loop__rb2z}
\eeq
and at the upper end, 
\beq
\xi_{IR,3\ell} \to 0 \quad {\rm as} \ \ r \nearrow r_{b1z}=\frac{11}{2} \ . 
\label{xiir_3loop_rb1z}
\eeq
The ratio $\xi_{IR,3\ell}/\xi_{IR,2\ell}$ increases
monotonically from 0 as $r$ increases from the value $r=34/13$ at the lower end
of the interval $I_r$, and this ratio approaches 1 from below as $r$ approaches
the upper end of the interval $I_r$ at $r=11/2$.  It is useful here and below
to give illustrative values of various quantities and ratios at an illustrative
value of $r$. For this purpose, we choose an $r$ approximately in the middle of
the $I_r$, namely $r=4$.  We have 
\beq
\xi_{IR,2\ell}{}|_{r=4} = \frac{2\pi}{3} = 2.0944
\label{xiir_2loop_req4}
\eeq
and
\beq
\xi_{IR,3\ell}{}|_{r=4} = \frac{4\pi(-2+\sqrt{22} \ )}{27} = 1.2522
\label{xiir_3loop_req4}
\eeq
so that
\beq
\frac{\xi_{IR,3\ell}}{\xi_{IR,2\ell}}{}\Big |_{r=4} = 
\frac{2(-2+\sqrt{22} \ )}{9} = 0.5979 \ . 
\label{xiir_23loopratio_req4}
\eeq
This ratio provides an illustrative measure of the decrease in the value of the
IR zero of $\beta$ when one calculates it at three-loop order, as compared with
two-loop order. 

% =====================================================================

\subsection{Four-Loop Level} 

At the four-loop level, the IR zero of $\beta_\xi$ is the (smallest positive)
root of the cubic equation
\beq
\beta_{\xi,4\ell,r} \equiv 
\tilde b_1 + \tilde b_2 \xi + \tilde b_3 \xi^2 + \tilde b_4 \xi^3 = 0 \ . 
\label{betaxi_4loop_reduced}
\eeq
Now $\tilde b_2 < 0$ for $r \in I_r$, and we recall our discussion above, 
that $\tilde b_3 < 0$ for $r \in I_r$ in the $\overline{MS}$ scheme and other 
schemes that maintain the existence of the IR zero in $\beta_{\xi,2\ell}$ at 
the three-loop level.  We can therefore write
Eq. (\ref{betaxi_4loop_reduced}) as 
\beq
\tilde b_1 - |\tilde b_2| \xi - |\tilde b_3| \xi^2 + \tilde b_4 \xi^3 = 0 \ . 
\label{beta_x_4loop_reduced_mag}
\eeq
For $r \in I_r$, Eq. (\ref{betaxi_4loop_reduced}), or equivalently,
(\ref{beta_x_4loop_reduced_mag}), has three real roots, and from these we
determine the relevant (smallest, positive) one as $\xi_{IR,4\ell}$. 
We list values of $\xi_{IR,4\ell}$ in Table \ref{xiir_nloop_values}.  

% =====================================================================

\subsection{Shift of IR Zero From $n$-Loop to $(n+1)$-Loop Level} 

In \cite{bc}, a general result was derived concerning the sign of the shift of
the IR zero of $\beta$ going from the $n$-loop level to the $(n+1)$-loop
level. Provided the scheme has the property that $b_\ell$ with $\ell \ge 3$ 
are such as to maintain the existence of the zero in the two-loop $\beta$
function, then $\alpha_{IR,(n+1)\ell} > \alpha_{IR,n\ell}$ if
if $b_{n+1} > 0$ and $\alpha_{IR,(n+1)\ell} < \alpha_{IR,n\ell}$ if
if $b_{n+1} < 0$.  The same proof can be applied here to deduce that, 
provided that the scheme has the property that $\hat b_\ell$ with $\ell \ge 3$ 
are such as to maintain the existence of the zero in the two-loop $\beta_\xi$
function, then
\beqs
& & \xi_{IR,(n+1)\ell} > \xi_{IR,n\ell} \quad {\rm if} \ \ 
\hat b_{n+1} > 0, \cr\cr
& & 
\xi_{IR,(n+1)\ell} < \alpha_{IR,n\ell}  \quad {\rm if} \ \ 
\hat b_{n+1} < 0  \ . 
\cr\cr
& & 
\label{xiir_nnp1_inequality}
\eeqs
We may apply this inequality for the comparisons of $\xi_{IR,3\ell}$ with 
$\xi_{IR,2\ell}$ and $\xi_{IR,4\ell}$ with $\xi_{IR,3\ell}$. Since 
$\hat b_3 < 0$ for $r \in I_r$, this result provides another way of
deducing the inequality (\ref{xiir_2versus3loop}) for the two-loop versus
three-loop comparison.  Applying the general inequality for the three-loop 
versus four-loop comparison, we infer that 
\beqs
& & \xi_{IR,4\ell} < \xi_{IR,3\ell}  \quad {\rm if} \ \ 2.615 < r < 3.119,
\ {\rm so} \ \tilde b_4 < 0, \cr\cr
& & 
\xi_{IR,4\ell} > \xi_{IR,3\ell}  \quad {\rm if} \ \ 3.119 < r < 5.500, 
\ {\rm so} \ \tilde b_4 > 0 \ . \cr\cr
& & 
\label{xiir_34loop_comparison}
\eeqs
These inequalities are evident in Table \ref{xiir_nloop_values}.  For example,
at $r=3.0$, $\xi_{IR,4\ell}/\xi_{IR,3\ell}=0.976$, while for $r=5.0$,
$\xi_{IR,4\ell}/\xi_{IR,3\ell}=1.02$.  One sees that the magnitude of the
fractional difference
\beq
\frac{|\xi_{IR,4\ell} - \xi_{IR,3\ell}|}{\xi_{IR,4\ell}}
\label{xiir_34loop_fracdif}
\eeq
is reasonably small.  This is in agreement with one's general expectation that
if a perturbative calculation is reliable, then as one calculates this quantity
to progressively higher-loop order, the magnitudes of the fractional
differences between the values at the $n$'th and $(n+1)$'th orders should
decrease.

% =====================================================================

\subsection{Summary of Results on IR Zero of $\beta_\xi$} 

We summarize our findings concerning $\xi_{IR,n\ell}$ as follows. As one goes
from the (scheme-independent) two-loop level to the three-loop level, the value
of the IR zero of $\beta_\xi$ decreases.  For $r$ in the lower part of the
interval $I_r$ where the two-loop $\beta_\xi$ function has an IR zero, this
reduction in the value of the IR zero is rather substantial.  For example, for
$r=3.0$, near the lower end of the interval $I_r$,
$\xi_{IR,3\ell}/\xi_{IR,2\ell} = 0.234$, while for $r=5.0$, near the upper
end of $I_r$, $\xi_{IR,3\ell}/\xi_{IR,2\ell}=0.873$. Going from three-loop to
four-loop order, the change in the value of the IR zero is smaller in magnitude
and can be of either sign, depending on the value of $r \in I_r$.  In general, 
both $\xi_{IR,3\ell}$ and $\xi_{IR,4\ell}$ are smaller than
$\xi_{IR,2\ell}$. 

% =======================================================================

\section{Some Structural Properties of $\beta_\xi$}
\label{structural}

For theories which exhibit an IR zero, $\xi_{IR,2\ell}$, in the two-loop beta
function, $\beta_{\xi,2\ell}$, there are several structural properties of
interest in addition to higher-loop values of this IR zero.  These include
\begin{itemize}

\item the value of $\xi$ at which $\beta_{\xi,n\ell}$ reaches a minimum 
in the interval $I_\xi$, denoted $\xi_{m,n\ell}$, where the subscript $m$
denotes minimum 

\item the value of
$\beta_{\xi,n\ell}$ at this minimum, denoted $(\beta_{\xi,n\ell})_{min}$

\item the derivative of $\beta_{\xi,n\ell}$ at $\xi_{IR,n\ell}$, denoted 
\beq
\beta'_{\xi,IR,n\ell} \equiv \frac{d\beta_{\xi,n\ell}}{d\xi}{}\Big |_{\xi = 
\xi_{IR,n\ell}} \ . 
\label{betaxiprime}
\eeq
\end{itemize}
Note that because $\xi = \alpha N_c$ and 
$\beta_\xi = \lim_{LNN} \beta_{\alpha}N_c$, the factor of $N_c$ divides
out in the derivative $d\beta_\xi/d\xi$, so that 
\beq
\frac{d\beta_\xi}{d\xi} = \lim_{LNN} \frac{d\beta_\alpha}{d\alpha} \ , 
\label{betaxiprime_betaprime}
\eeq
and
\beq
\frac{d\beta_{\xi,n\ell}}{d\xi} = \lim_{LNN} 
\frac{d\beta_{\alpha,n\ell}}{d\alpha} \ , 
\label{betaxiprime_betaprime_nloop}
\eeq
In particular, 
\beq
\frac{d\beta_{\xi,n\ell}}{d\xi}{}\Big |_{\xi = \xi_{IR,n\ell}} = 
\lim_{LNN} 
\frac{d\beta_{\alpha,n\ell}}{d\alpha}{}\Big |_{\alpha = \alpha_{IR,n\ell}} \ .
\label{betaxiprime_betaprime_nloop_rel}
\eeq
As was discussed in \cite{bc}, higher-loop calculations of the derivative
$\frac{d\beta_{\alpha,n\ell}}{d\alpha} {}|_{\alpha = \alpha_{IR,n\ell}}$ are of
interest because this enters into estimates of a dilaton mass in a
quasiconformal gauge theory.  In turn, this also provides one motivation for
studying the LNN limit of this derivative,
$\frac{d\beta_{\xi,n\ell}}{d\xi}{}|_{\xi = \xi_{IR,n\ell}}$.

% ====================================================================

\subsection{Position of Minimum in $\beta_{\xi,n\ell}$} 

Concerning the position of the minimum in $\beta_\xi$ for $r \in I_r$, we 
calculate that at the two-loop level, 
\beq
\xi_{m,2\ell} = \frac{8\pi(11-2r)}{3(13r-34)} \ . 
\label{xi_critical_2loop}
\eeq
This satisfies 
\beq
\xi_{m,2\ell} = \frac{2}{3}\xi_{IR,2\ell} \ . 
\label{xi_critical_23loop_ratio}
\eeq
At the three-loop level in the $\overline{MS}$ scheme, we find 
\beq
\xi_{m,3\ell} = 
\frac{3\pi[ -9(13r-34)+\sqrt{E_{3\ell}} \ ]}{D_{3\ell}} \ ,
\label{xi_critical_3loop}
\eeq
where we define the shorthand notation 
\beq
E_{3\ell} = -409196+320604r-60711r^2+3584r^3 \ . 
\label{e3ell}
\eeq
From Eqs. (\ref{xiir_3loop_explicit}) and (\ref{xi_critical_3loop}), we find
\beq
\frac{\xi_{m,3\ell}}
     {\xi_{IR,3\ell}} =
\frac{-9(13r-34)+\sqrt{E_{3\ell}}}{4[-3(13r-34)+\sqrt{C_{3\ell}} \ ]} \ . 
\label{xi_critical_3loop_over_xiir_3loop}
\eeq
This may be compared with the corresponding ratio of two-loop quantities
$\xi_{m,2\ell}/\xi_{IR,2\ell}=2/3$.  In contrast to the latter ratio,
which is a constant, independent of $r \in I_r$, the ratio
(\ref{xi_critical_3loop_over_xiir_3loop}) is a monotonically decreasing
function of $r \in I_r$.  At the lower end of this interval, 
\beq
\frac{\xi_{m,3\ell}}
     {\xi_{IR,3\ell}} = \frac{1}{\sqrt{2}} \quad 
{\rm at} \ \ r=r_{b2z} = \frac{34}{13} \ . 
\label{ximinir_rat_3loop_rb2z}
\eeq
As $r$ approaches the upper end of the $I_r$ at $r_{b1z}=11/2$, the ratio
(\ref{xi_critical_3loop_over_xiir_3loop}) approaches the limit 
\beq
\lim_{r \nearrow r_{b1z}} \frac{\xi_{m,3\ell}}{\xi_{IR,3\ell}} = \frac{2}{3} 
 \ . 
\label{ximinir_rat_3loop_rb1z}
\eeq
Note that both $\xi_{m,3\ell}$ and $\xi_{IR,3\ell}$ individually approach 
zero as $r \nearrow 11/2$, although their ratio in
Eq. (\ref{ximinir_rat_3loop_rb1z}) approaches a constant.

Illustrative values in the LNN limit for $r=4$ are
\beq
\xi_{m,2\ell}{}|_{r=4} = \frac{4\pi}{9} = 1.396.
\label{xi_critical_2loop_req4}
\eeq
and
\beq
\xi_{m,3\ell}{}|_{r=4} = \frac{2\pi(-1+\sqrt{5} \ )}{9} = 0.8629.
\label{xi_critical_3loop_lnn_req4}
\eeq
so that for this value, $r=4$, in addition to the ratio
$\xi_{m,2\ell}/\xi_{IR,2\ell} = 2/3$, we have
\beq
\frac{\xi_{m,3\ell}}{\xi_{IR,3\ell}}{}\Big |_{r=4} = 
\frac{3}{2} \Big ( \frac{-1+\sqrt{5}}{-2+\sqrt{22}} \Big ) = 0.68915 \ . 
\label{xi_critical_3loop_req4}
\eeq
and
\beq
\frac{\xi_{m,3\ell}}
     {\xi_{m,2\ell}} = \frac{-1 + \sqrt{5}}{2} = 0.6180.
\label{xi_critical_2over3loop_req4}
\eeq
%

% =======================================================================

\subsection{Value of $\beta_{\xi,n\ell}$ at Minimum}

We calculate the following minimum values of $\beta_{\xi,n\ell}$ as a function
of $r \in I_r$: 
\beq
(\beta_{\xi,2\ell})_{min} = -\frac{2^5\pi(11-2r)^3}{3^4(13r-34)^2}
\label{betaxi_2loop_min}
\eeq
and
\beq
(\beta_{\xi,3\ell})_{min} = - \frac{3\pi(F_{3\ell} +
G_{3\ell}\sqrt{E_{3\ell}} \ )}{8D_{3\ell}^3} \ , 
\label{betaxi_3loop_min}
\eeq
where $D_{3\ell}$ and $E_{3\ell}$ were defined above in Eqs. 
(\ref{d3ell}) and (\ref{e3ell}).  The functions $F_{3\ell}$ and $G_{3\ell}$ are
given in the appendix. 

For the illustrative value $r=4$,
\beq
(\beta_{\xi,2\ell})_{min} = -\frac{2^3\pi}{3^5} = -0.1034
\label{betaxi_2loop_min_req4}
\eeq
and
\beq
(\beta_{\xi,3\ell})_{min} = - \frac{2\pi(13-5\sqrt{5} \ )}{3^5} = -0.04705 \ . 
\label{betaxi_3loop_min_req4}
\eeq
%

% ===================================================================

\subsection{ $d\beta_{\xi,n\ell}/d\xi$ at $\xi_{IR,n\ell}$} 

At the two-loop level, we calculate 
\beq
\beta_{\xi,IR,2\ell}' = \frac{2(11-2r)^2}{3(13r-34)} \ . 
\label{dbetaxi_2loop_dxi_at_xiir_2loop}
\eeq
This is clearly positive for $r \in I_r$, approaching zero as $r$ approaches
the upper end of this interval at $r=r_{b1z}$.  At the three-loop level, we
find 
\beq
\beta_{\xi,IR,3\ell}' = 
\frac{4[-3(13r-34)C_{3\ell}+K_{3\ell}\sqrt{C_{3\ell}} \ ]}{D_{3\ell}^2} \ ,
\label{dbetaxi_3loop_dxi_at_xiir_3loop}
\eeq
where $C_{3\ell}$ and $D_{3\ell}$ were defined above in Eqs. (\ref{c3ell}) and
(\ref{d3ell}), and we define 
\beq
K_{3\ell} = -21023+16557r-3129r^2+224r^3 \ . 
\label{edb_3loop}
\eeq

The first term in the numerator, $-3(13r-34)C_{3\ell}$, is negative for $r \in
I_r$, but is smaller in magnitude than the second term,
$K_{3\ell}\sqrt{C_{3\ell}}$.  This shows analytically that
$\beta_{\xi,IR,3\ell}' > 0$ for $r \in I_r$. The positivity of
$\beta_{\xi,IR,n\ell}'$ for $r \in I_r$ is obvious from the graph of
$\beta_{\xi,n\ell}$. Since this function is continuous, is negative for $0 <
\xi < \xi_{IR,n\ell}$, and has, generically, a simple zero at
$\beta_{\xi,n\ell}' > 0$, it follows that $\beta_{IR,n\ell}' > 0$ for $r \in
I_r$.  We have also calculated $\beta_{\xi,IR,4\ell}'$ analytically, but the
expression is somewhat cumbersome, since it involves cube roots, so we do not
list it.  Illustrative results of these ratios for the LNN limit and the
typical value, $r=4$, are
\beq
\beta_{\xi,IR,2\ell}'{}|_{r=4} = \frac{1}{3} \ ,
\label{dbetaxi_2loop_dxi_at_xiir_2loop_req4}
\eeq
and
\beq
\beta_{\xi,IR,3\ell}'{}|_{r=4} = \frac{4(-44+13\sqrt{22} \ )}{3^5} = 0.2794 \ .
\label{dbetaxi_3loop_dxi_at_xiir_3loop_req4}
\eeq
This illustrates how the slope of the $\beta_\xi$ function at the $n$-loop IR
zero decreases as it is calculated to three-loop order, as compared with the
2-loop calculation.  This is also shown by the explicit numerical results in
the tables going up to four-loop order. 

% =====================================================================

\section{Anomalous Dimension $\gamma_m$}
\label{anomdim}

The anomalous dimension $\gamma_m$ describes the scaling of a
fermion bilinear and the running of a dynamically generated fermion mass in the
phase with spontaneous chiral symmetry breaking.  $\gamma_m$ is defined as
$\gamma_m = d\ln Z_m/dt$, where $Z_m$ is the corresponding renormalization
constant for the fermion bilinear operator.  In the non-Abelian Coulomb phase
(which is a conformal phase), the IR zero of $\beta$ is exact, although a
calculation of it to a finite-order in perturbation theory is only approximate,
and $\gamma_m$ evaluated at this IR fixed point is exact. In the phase with
spontaneous chiral symmetry breaking, where an IR fixed point, if it exists, is
only approximate, $\gamma_m$ is an effective quantity describing the running of
a dynamically generated fermion mass for the evolution of the theory near this
approximate IRFP. As in \cite{bvh,bc}, for notational simplicity we will often
suppress the subscript $m$ on $\gamma_m$ where the meaning is clear. 

This anomalous dimension can be expressed as a series in $a$
or equivalently, $\alpha$:
\beq
\gamma \equiv \gamma_m = \sum_{\ell=1}^\infty c_\ell \, a^\ell
         = \sum_{\ell=1}^\infty \bar c_\ell \, \alpha^\ell \ ,
\label{gamma}
\eeq
where $\ell$ denotes the loop order and $\bar c_\ell = c_\ell/(4\pi)^\ell$ is
the $\ell$-loop series coefficient.  The coefficient $c_1$ is
scheme-independent while the $c_\ell$ with $\ell \ge 2$ are
scheme-dependent. The $c_\ell$ have been calculated up to $\ell=4$ in the
$\overline{MS}$ scheme \cite{gamma4}.  The $n$-loop expression for $\gamma$ is
denoted $\gamma_{n\ell}$ and is given by Eq. (\ref{gamma}) with $\infty$
replaced by $n$ as the upper limit on the summation over $\ell$.  In \cite{bvh}
we evaluated $\gamma$ to three- and four-loop order at the IR zero of $\beta$
calculated to the same order and showed that these higher-loop results were
somewhat smaller than the two-loop evaluation.  In \cite{sch} we discussed
scheme transformations at an IR fixed point and implications, including those
for the scheme-dependence of $\gamma$.

In the IR-conformal phase, unitarity implies a lower limit on the dimension of
a spinless operator ${\cal O}$, namely, $D_{\cal O} \ge (d-2)/2$, where $d$ is
the spacetime dimension \cite{cftbound}. With our sign convention,
\beq
D_{\bar\psi\psi} = 3-\gamma \ ,
\label{dgam}
\eeq
so that in this IR-conformal phase, $\gamma$ is bounded above as
\beq
\gamma \le 2 \ .
\label{gammabound}
\eeq
In the IR phase with confinement and spontaneous chiral symmetry breaking, the
dynamical mass generated for the fermions behaves, for Euclidean momentum large
compared with the chiral-symmetry-breaking scale, $\Lambda$, as $\Sigma(k)
\propto \Lambda (\Lambda/k)^{2-\gamma}$ (up to a logarithmic factor) for $k
>> \Lambda$.  Hence, the upper bound (\ref{gammabound}) also applies in this
phase.

For our present analysis of the LNN limit, we reexpress $\gamma$ in terms of
$x$ or equivalently, $\xi$: 
\beq
\gamma = \sum_{\ell=1}^\infty \hat c_\ell \, x^\ell
         =  \sum_{\ell=1}^\infty \tilde c_\ell \, \xi^\ell \ ,
\label{gammax}
\eeq
where
\beq
\hat c_\ell = \lim_{LNN} \frac{c_\ell}{N_c^\ell}
\label{chat_ell}
\eeq
and
\beq
\tilde c_\ell = \lim_{LNN} \frac{\bar c_\ell}{N_c^\ell} \ . 
\label{ctilde_ell}
\eeq
Thus, similarly to the relation between $\bar c_\ell$ and $c_\ell$,
\beq
\tilde c_\ell = \frac{\hat c_\ell}{(4\pi)^\ell} \ . 
\label{tildehatcell}
\eeq
The $n$-loop expression for $\gamma$ is given by the right-hand side
of Eq. (\ref{gammax}) with the sum running from $\ell=1$ to $\ell=n$ rather
than $\ell=\infty$.

In the LNN limit we find 
\beq
\hat c_1 = 3 
\label{chat1}
\eeq
\beq
\hat c_2 = \frac{203}{12} - \frac{5}{3} r 
\label{chat2}
\eeq
\beqs
\hat c_3 & = & \frac{11413}{108}-\bigg ( \frac{1177}{54} + 12\zeta(3) \bigg ) r
- \frac{35}{27}r^2 
\cr\cr
& = & 105.676 - 36.221 r - 1.296 r^2 
\label{chat3}
\eeqs
and
\begin{widetext}
\beqs
\hat c_4 & = & \frac{460151}{576}-\frac{23816}{81}r+\frac{899}{162}r^2
-\frac{83}{81}r^3 + 
\bigg ( \frac{1157}{9}-\frac{889}{3}r+20r^2+\frac{16}{9}r^3 \bigg ) \zeta(3) 
\cr\cr & + & r\Big (66-12r \Big ) \zeta(4)+ 
              \Big (-220+160r \Big ) \zeta(5) \cr\cr
& = & 725.280 -412.892r + 16.603r^2 + 1.1123 r^3 \ , 
\label{chat4}
\eeqs
\end{widetext}
where the floating-point numerical results are given to the indicated 
accuracy. We list numerical values of the corresponding coefficients 
$\tilde c_\ell$ for $1 \le \ell \le 4$ in Table \ref{ctilde_nloop_values}. 

In \cite{bvh}, the value of $\gamma_{n\ell}$ evaluated
at $\alpha=\alpha_{IR,n\ell}$ was denoted as $\gamma_{_{IR,n\ell}}$  and, 
analogously, here, in the LNN limit, we define
\beq
\gamma_{_{IR,n\ell}} \equiv \gamma_{n\ell}{}\Big |_{\xi=\xi_{IR,n\ell}} \ .
\label{gamma_xi_ir_nloop}
\eeq

At the two-loop level, in terms of the coefficients of $\beta$ and 
$\gamma_m$, taking into account that $\hat b_2 < 0$ for $r \in I_r$, 
\beq
\gamma_{_{IR,2\ell}} = \frac{\hat b_1(\hat c_1 |\hat b_2| + \hat c_2 \hat b_1)}
{\hat b_2^2} \ . 
\label{gamma_ir_2loop_bcform}
\eeq
Note that the sum of the $\ell$-values of the products of coefficients in the 
denominator of the expression for $\gamma_{_{IR,2\ell}}$ is equal to the sum of
the $\ell$-values of the coefficients in the numerator.  In this case, this sum
is 4.  Because of this homogeneity property and the relations in
Eqs. (\ref{tildehatbell}) and (\ref{tildehatcell}), it follows that 
$\gamma_{_{IR,2\ell}}$ has the same form as Eq. (\ref{gamma_ir_2loop_bcform})
if one replaces each $\hat b_\ell$ by $\tilde b_\ell$ and each $\hat c_\ell$ by
$\tilde c_\ell$. Explicitly \cite{bvh}, 
\beq
\gamma_{_{IR,2\ell}} = \frac{(11-2r)(1009-158r+40r^2)}{12(13r-34)^2} \ .
\label{gamma_irfp_2loop_fund_lnn}
\eeq
This $\gamma_{_{IR,2\ell}}$ decreases monotonically as $r$ increases from
$=r_{b2z}=34/13$ at the lower end of the interval $I_r$ to zero at
$r=r_{b1z}=11/2$ at the upper end of this interval.  Because of the upper bound
(\ref{gammabound}), if the value of $\gamma$ calculated via this truncated
perturbative expansion is greater than 2, then it is unphysical. This applies,
in particular, to the divergence in $\gamma_{_{IR,2\ell}}$ at $r=r_{b2z}$.  We
find that in the relevant interval $I_r$, $\gamma_{_{IR,2\ell}}$ exceeds 2 as
$r$ decreases below the value 3.569.  Thus, we cannot use the formula
(\ref{gamma_irfp_2loop_fund_lnn}) for $r < 3.569$, and it is subject to large
corrections unless $r$ is substantially above this value. For reference, we
find that $\gamma_{IR,2\ell}$ exceeds 1 as $r$ decreases below the value 
$r=3.879$.

At the three-loop level, 
\beq
\gamma_{_{IR,3\ell}} = x( \hat c_1 + \hat c_2 x + \hat c_3 x^2){}\Big
|_{x=x_{IR,3\ell}} \ . 
\label{gamma_ir_3loop}
\eeq
Substituting $x_{IR,3\ell}=\xi_{_{IR,3\ell}}/(4\pi)$ from
Eq. (\ref{xiir_3loop}) and using the property that $\hat b_2 < 0$ for $r \in
I_r$ and, as discussed above, the property that $\hat b_3 < 0$ in a scheme that
preserves the existence of the IR zero in the (scheme-independent)
$\beta_{\xi,2\ell}$ at the three-loop level, we can write $\gamma_{IR,3\ell}$
in terms of positive quantities as
\begin{widetext}
\beqs
& & \gamma_{_{IR,3\ell}} = \frac{1}{4|\hat b_3|^3}
\bigg [ -|\hat b_2| + \sqrt{\hat b_2^2 + 4 \hat b_1|\hat b_3|} \ \bigg ] 
\bigg [ 2\hat c_1 |\hat b_3|^2 + \hat c_2 |\hat b_2||\hat b_3| 
- \hat c_3(\hat b_2^2+2\hat b_1 |\hat b_3|) 
+ (-\hat c_2 |\hat b_3|+\hat c_3 |\hat b_2|)\sqrt{\hat b_2^2 +
4\hat b_1 |\hat b_3|} \ \bigg ] \ . \cr\cr
& & 
\label{gamma_ir_3loop_bc}
\eeqs
\end{widetext}
Thus, $\gamma_{_{IR,3\ell}}$ has the same type of homogeneity property as a
function of the $\hat b_\ell$ and $\hat c_\ell$ coefficients as we discussed
for $\gamma_{_{IR,2\ell}}$.  Hence, $\gamma_{_{IR,3\ell}}$ is identically expressed
by replacing each $\hat b_\ell$ by $\tilde b_\ell$ and each $\hat c_\ell$ by
$\tilde c_\ell$ in Eq. (\ref{gamma_ir_3loop_bc}).  Inserting the explicit
expressions for the $\hat b_\ell$ and $\hat c_\ell$ then yields the explicit
result as a function of $r$.  For the present work we have also calculated the
four-loop anomalous dimensions evaluated at the IR zero of $\beta_\xi$
evaluated to the same order, $\gamma_{_{IR,4\ell}}$. 

We list numerical values of these quantities in Table
\ref{gamma_ir_nloop_values}.  We find that, as was the case with
$\gamma_{_{IR,2\ell}}$, \ $\gamma_{_{IR,3\ell}}$ decreases monotonically as $r$
increases from $=r_{b2z}$ at the lower end of the interval $I_r$ to zero at
$r=r_{b1z}$ at the upper end of this interval.  At $r=r_{b2z}$,
$\gamma_{_{IR,3\ell}} = 2.680$, and hence the value of $\xi_{_{IR,3\ell}}$ is
evidently too large for the (perturbative) calculation of this anomalous
dimension to be reliable.  $\gamma_{_{IR,3\ell}}$ decreases through the value 2
as $r$ increases through the value 2.73, and $\gamma_{_{IR,3\ell}}$ decreases
further through the value 1 as $r$ increases through 3.09. Just as was true of
particular values of $N_c$ and $N_f$ \cite{bvh}, here, for a given value of
$r$,
\beq
\gamma_{_{IR,3\ell}} \le \gamma_{_{IR,2\ell}} \ , 
\label{gamma_23loop_rel}
\eeq
with equality only at $r=r_{b1z}$, where 
$\gamma_{_{IR,3\ell}} =  \gamma_{_{IR,2\ell}}=0$. 

% =====================================================================

\section{Approach to the LNN Limit}
\label{approach}

One of the motivations for the present work on calculations of the IR zero, the
various structural properties of $\beta$, and the anomalous dimension $\gamma$
evaluated at the IR zero in the LNN limit is that the results provide an
understanding of (i) the approximate universality that is exhibited by
calculations of these quantities for theories with different values of $N_c$
and $N_f$ such that the respective values of $r$ are the same or nearly the
same, and (ii) the fact that this approximate universality occurs even for
moderate values of $N_c$ and $N_f$.  As we have shown above, the quantities 
\begin{itemize}

\item $\alpha_{IR,n\ell} N_c$ 

\item $\alpha_{m,n\ell} N_c$ 

\item $(\beta_{\alpha,n\ell})_{min} N_c$ 

\item $\frac{d\beta_{IR,n\ell}}{d\alpha} {}\Big |_{\alpha=\alpha_{IR,n\ell}}$ 

\item $\gamma_{n\ell} {}\Big |_{\alpha=\alpha_{IR,n\ell}}$ 

\end{itemize}
are finite in the LNN limit, and we will next show that they approach their 
respective LNN limiting values, namely 
\begin{itemize}

\item $\xi_{IR,n\ell}$ 

\item $\xi_{m,n\ell}$

\item $(\beta_{\xi,n\ell})_{min}$ 

\item $\frac{d\beta_{\xi,n\ell}}{d\xi} {}\Big |_{\xi=\xi_{IR,n\ell}}$

\item $\gamma_{n\ell}{}\Big |_{\xi=\xi_{IR,n\ell}}$

\end{itemize}
rather rapidly as $N_c$ increases (with $r=N_f/N_c$ fixed). The reason for this
is that the largest correction terms to the respective LNN limits for these
quantities are of order $1/N_c^2$ and hence are rather strongly suppressed even
for moderate values of $N_c$ and $N_f$.  In turn, this is a consequence of the
$N_c$-dependence of the relevant group invariants for fermions in the
fundamental representation, in particular, the Casimir invariant
$C_f=(N_c^2-1)/(2N_c)$ \cite{casimir}.

We first exhibit the structure of the correction terms for the coefficients in
the beta function and anomalous dimension $\gamma$.  Multiplying by the
appropriate inverse powers of $N_c$ to obtain finite results in the LNN limit,
we thus consider $b_\ell/N_c^\ell$ and $c_\ell/N_c^\ell$.  These are
polynomials in $r$ and in the variable $1/N_c^2$.  We find the following
explicit results for correction terms:
\beq
\frac{b_1}{N_c} = \hat b_1   
\label{b1_lnn}
\eeq
\beq
\frac{b_2}{N_c^2} = \hat b_2 + \frac{r}{N_c^2} 
\label{b2_lnn}
\eeq
\beqs
& & \frac{b_3}{N_c^3} = \hat b_3 +
\frac{11r(17-2r)}{36N_c^2} + \frac{r}{4N_c^4} \cr\cr
& & 
\label{b3_lnn}
\eeqs
and
\begin{widetext}
\beqs 
\frac{b_4}{N_c^4} & = & \hat b_4 + \frac{1}{N_c^2}\bigg [
-\frac{40}{3}+\frac{58583r}{1944}-\frac{2477r^2}{243}-\frac{77r^3}{243}
+ \Big ( 352-\frac{548r}{9} - \frac{64r^2}{9} \Big )\zeta(3) \bigg ] \cr\cr
& + & \frac{r}{N_c^4}\bigg [ -\frac{2341}{216} - \frac{623r}{54} + 
\frac{4}{9}(11+61r)\zeta(3) \bigg ] - \frac{23r}{8N_c^6} \ . 
\label{b4fund_lnn}
\eeqs
\end{widetext}
For the ratios $c_\ell/N_c^\ell$ we find 
\beq
\frac{c_1}{N_c} = \hat c_1 - \frac{3}{N_c^2} 
\label{c1_lnn}
\eeq
\beq
\frac{c_2}{N_c^2} = \hat c_2 + \frac{-53+5r}{3N_c^2} + \frac{3}{4N_c^4}
\label{c2_lnn}
\eeq
and
\beqs 
& & \frac{c_3}{N_c^3}= \hat c_3 + \frac{1}{N_c^2}\Big (
-\frac{26309}{216} + \frac{899}{27}r + \frac{35}{27}r^2 \Big ) \cr\cr & + &
\frac{1}{N_c^4} \Big ( \frac{129}{4} - \frac{23r}{2} + 12\zeta(3)r \Big ) -
\frac{129}{8N_c^6} \ .
\label{c3_lnn}
\eeqs
The corresponding result for $c_4$ is given in the appendix.  As noted, these
expansions exhibit the feature that the largest subleading term is smaller than
the leading term by a factor of $1/N_c^2$ in the LNN limit.

Similarly, the approach of the product $\alpha_{IR,2\ell}N_c$ to its LNN limit,
$\xi_{IR,2\ell}$, has the form 
\beq
\alpha_{IR,2\ell}N_c = \frac{4\pi(11-2r)}{13r-34}
+ \frac{12\pi r(11-2r)}{(34-13r)^2N_c^2} + O \Big ( \frac{1}{N_c^4} \Big ) \ , 
\label{alfir_2loop_lnn}
\eeq
where the first term on the right-hand side is the LNN limit, $\xi_{IR,2\ell}$,
given in Eq. (\ref{xiir_2loop}).  Similarly, the two-loop anomalous dimension
$\gamma_m$, evaluated at the IR zero of the two-loop $\beta$ function, has the
expansion
\beqs
& & \gamma_{_{IR,2\ell}} = \frac{(11-2r)(1009-158r+40r^2)}{12(13r-34)^2} \cr\cr
& + & \frac{(11-2r)(18836-5331r+648r^2-140r^3)}{(13r-34)^3 N_c^2} + 
O \Big ( \frac{1}{N_c^4} \Big ) \ . \cr\cr
& & 
\label{gamma_ir_2loop_lnn} 
\eeqs
where the first term on the right-hand side is the LNN limit, given in Eq.
(\ref{gamma_irfp_2loop_fund_lnn}).  Similar results hold for the expressions
calculated to higher-loop orders. Analogous results concerning correction
terms to LNN limits apply to the theory with ${\cal N}=1$ supersymmetry, as
will be clear from our discussion below.

We next give some numerical results illustrating the rapidity of approach to 
the LNN limit. For definiteness, we consider two cases: $N_c=3$, $N_f=12$, so
$r=4$, and $N_c=3$, $N_f=15$, so $r=5$. For the first case, $N_c=3$, $N_f=12$,
the two-loop, three-loop, and four-loop values of the IR zero of $\beta$ are
\beq
\alpha_{IR,2\ell} = 0.7540, \quad \alpha_{IR,3\ell} = 0.4349, \quad
\alpha_{IR,4\ell} = 0.4704 \ . 
\label{alfir_nloop_nc3nf12}
\eeq
(In Table III of \cite{bvh} these were listed these to three significant
figures; here we list them to higher accuracy to compare with the estimates
from the LNN limit.)  To compute the LNN approximations to
$\alpha_{IR,n\ell}$ calculated for a specific $N_c$ and $N_f=rN_c$, we take 
the corresponding values for $\xi_{IR,n\ell}$ from
Table \ref{xiir_nloop_values} and divide by $N_c$, obtaining the result
\beq
\alpha_{IR,n\ell,L} \equiv \frac{\xi_{IR,n\ell}}{N_c} \quad {\rm for \ fixed} \
r \ , 
\label{alfir_nloop_flnn}
\eeq
where the subscript $L$ indicates the LNN origin of the estimate.  These
approximations become progressively more accurate as $N_c$ gets large for fixed
$r$, but as we will show, they are already quite close to the actual values in
Eq. (\ref{alfir_nloop_nc3nf12}).  We obtain
\beq
\alpha_{IR,2\ell,L} = 0.6982, \quad \alpha_{IR,3\ell,L} = 0.4713, \quad
\alpha_{IR,4\ell,L} = 0.4497
\label{alfir_nloop_flnn_nc3nf12}
\eeq
Hence, for this $N_c=3$, $N_f=12$ case, we find 
\beq
\frac{\alpha_{IR,2\ell}}{\alpha_{IR,2\ell,L}} = 1.080, \quad 
\frac{\alpha_{IR,3\ell}}{\alpha_{IR,3\ell,L}} = 1.042, \quad 
\frac{\alpha_{IR,4\ell}}{\alpha_{IR,4\ell,L}} = 1.046, \quad
\label{alfir_nloop_ratios_nc3nf12}
\eeq
Carrying out the corresponding analysis for $N_c=3$,  $N_f=15$, we calculate
\beq
\frac{\alpha_{IR,2\ell}}{\alpha_{IR,2\ell,L}} = 1.057, \quad 
\frac{\alpha_{IR,3\ell}}{\alpha_{IR,3\ell,L}} = 1.046, \quad 
\frac{\alpha_{IR,4\ell}}{\alpha_{IR,4\ell,L}} = 1.046 \ . \quad 
\label{alfir_nloop_ratios_nc3nf15}
\eeq

We next perform the corresponding numerical comparisons for
$\gamma_{IR,n\ell}$ at the two-loop, three-loop, and four-loop levels. 
For $N_c=3$ and $N_f=12$, 
\beq
\gamma_{IR,2\ell} = 0.7728, \quad \gamma_{IR,3\ell} = 0.31175 \quad
\gamma_{IR,4\ell} = 0.2533 \ . 
\label{gamma_ir_nloop_nc3nf12}
\eeq
(In Table VI of \cite{bvh} these were listed to three significant figures; here
we list them to higher accuracy to compare with the estimates from the LNN
limit.) The LNN values of $\gamma_{IR,n\ell}$ are displayed in Table
\ref{gamma_ir_nloop_values}. Comparing these with the values in
Eq. (\ref{gamma_ir_nloop_nc3nf12}), we find, for this $N_c=3$, $N_f=12$ case,
the ratios
\beq
\frac{\gamma_{IR,2\ell}}{\gamma_{IR,2\ell,L}} = 0.985, \quad 
\frac{\gamma_{IR,3\ell}}{\gamma_{IR,3\ell,L}} = 0.913, \quad 
\frac{\gamma_{IR,4\ell}}{\gamma_{IR,4\ell,L}} = 0.880 \ . \quad 
\label{gamma_ir_nloop_ratios_nc3nf12}
\eeq
Performing the corresponding analysis for $N_c=3$,  $N_f=15$, 
we obtain 
\beq
\frac{\gamma_{IR,2\ell}}{\gamma_{IR,2\ell,L}} = 0.943, \quad 
\frac{\gamma_{IR,3\ell}}{\gamma_{IR,3\ell,L}} = 0.929, \quad 
\frac{\gamma_{IR,4\ell}}{\gamma_{IR,4\ell,L}} = 0.929 \ . \quad 
\label{gamma_ir_nloop_ratios_nc3nf15}
\eeq
These numerical comparisons show that even for a moderate value of $N_c$ such
as $N_c=3$, the values of these IR zeros of $\beta_{\alpha,n,\ell}$ and
anomalous dimensions $\gamma_{IR,n\ell}$ are close to the approximations that
one obtains from the LNN limit.  The agreement becomes better as $N_c$
increases. 

From our analysis, it follows that the approach to the LNN limit is of the same
form for other structural properties describing the UV to IR evolution of the
theory; that is, the first subleading correction term to the LNN limit is
suppressed by $1/N_c^2$.  We show this explicitly for additional two-loop
quantities.  The result for $\alpha_{m,2\ell}$ follows immediately from
Eqs. (\ref{xi_critical_2loop}), (\ref{xi_critical_23loop_ratio}), and
(\ref{alfir_2loop_lnn}). For $(\beta_{IR,2\ell})_{min}$, the approach to the
LNN limit has the form 
\beq
(\beta_{IR,2\ell})_{min} N_c = (\beta_{\xi,2\ell})_{min} - 
\frac{2^6 \pi r(11-2r)^3}{3^3(13r-34)^3 N_c^2} + O \Big ( \frac{1}{N_c^4} \Big
) \ , 
\label{beta_2loop_min_lnnapproach}
\eeq
where $(\beta_{\xi,2\ell})_{min}$ was given in Eq. (\ref{betaxi_2loop_min}). 
For the derivative of $\beta$ at the $n$-loop IR zero of $\beta$, the approach
to the LNN limit has the form 
\beq
\frac{d\beta_{\alpha,2\ell}}{d\alpha}{}\Big |_{\alpha=\alpha_{IR,2\ell}} = 
\frac{d\beta_{\xi,2\ell}}{d\xi}{}\Big |_{\xi=\xi_{IR,2\ell}} + 
\frac{2r(11-2r)^2}{(13r-34)^2N_c^2} + O \Big ( \frac{1}{N_c^4} \Big ) \ , 
\label{betaprime_2loop_lnnapproach}
\eeq
where $d\beta_{\xi,2\ell}/d\xi {}\Big |_{\xi=\xi_{IR,2\ell}} =
\beta'_{\xi,IR,2\ell}$ was given in Eq. 
(\ref{dbetaxi_2loop_dxi_at_xiir_2loop}). 

% =======================================================================

\section{Contrast with Fermions in Higher Representations} 
\label{contrast}

Gauge theories with fermions in higher-dimensional representations, in
particular, two-index representations, are also of interest \cite{sanrev}.  We
comment briefly here on the adjoint, and symmetric and antisymmetric rank-2
tensor representations with Young tableaux $\sym$ and $\asym$. Rather than an
LNN limit, in these cases, one takes $N_f$ equal to a (non-negative, integer)
constant as $N_c \to \infty$ to get a finite limit. For the adjoint
representation, $b_\ell/N_c^\ell$ is a constant for $\ell=1, \ 2, \ 3$, while
$b_4/N_c^4$ is equal to a constant plus a $1/N_c^2$ correction term (where
$b_\ell$ with $\ell=3, \ 4$ are calculated in the $\overline{MS}$ scheme). For
the S2 ($\sym$) and A2 ($\asym$) representations, symbolized together as T2,
the corrections to the $N_c \to \infty$ limit go like $1/N_c$ instead of
$1/N_c^2$.  For example,
\beq
\frac{b_1}{N_c} = \frac{11-2N_f}{3} \mp \frac{2N_f}{N_c}  \quad {\rm for}  \ 
T2 \ , 
\label{b1overnct2}
\eeq
where the upper (lower) sign applies to S2 (A2). 

These differences are reflected in the large-$N_c$ corrections to the $n$-loop
expressions for the IR zero of $\beta$. We illustrate this at the two-loop
level. For the adjoint representation, $N_{f,b1z}=11/4$ and $N_{f,b2z}=17/16$,
so that there is only a single integer value of $N_f$ for which the theory is
asymptotically free and has an IR zero in $\beta_{2\ell}$, namely $N_f=2$.
One has 
\beq
\alpha_{IR,2\ell}N_c = \frac{2\pi(11-4N_f)}{16N_f-17} \ , \quad R=adj 
\label{xiir_2loop_adj_lnn}
\eeq
(independent of $N_c$) so the right-hand side is equal to $2\pi/5$ for the
value $N_f=2$, independent of $N_c$.

For the S2 and A2 representations (denoted T2 again), for large $N_c$,
\beqs
\alpha_{IR,2\ell}N_c & = & \frac{2\pi(11-2N_f)}{8N_f-17} \pm 
\frac{6\pi N_f(-47+2N_f)}{(8N_f-17)^2N_c} \cr\cr
& + & O \Big ( \frac{1}{N_c^2} \Big ) \quad {\rm for} \ \ R = T2 \ , 
\label{xiir_2loop_t2_lnn}
\eeqs
so that in these S2 and A2 cases, the leading correction to the $N_c \to
\infty$ result goes like $1/N_c$.

\section{Supersymmetric Gauge Theory}
\label{susy}

% ========================================================================

\subsection{$\beta_{\xi,s}$ Function and IR Zeros} 

Here we consider the LNN limit of an asymptotically free, vectorial gauge
theory with ${\cal N}=1$ supersymmetry, gauge group $G={\rm SU}(N_c)$, and a
chiral superfield content $\Phi_i, \ \tilde \Phi_i$, $i=1,...,N_f$, in the
$\fund$, \ $\overline{\fund}$ representations, respectively. One of the appeals
of this theory is that a number of exact results on the infrared properties of
the theory are known \cite{nsvzbeta,seiberg}, so one can compare
perturbative predictions with these exact results.  This was done for general
$G$ and various representations $R$, $\bar R$ for the $N_f$ pairs of chiral
superfields $\Phi_i, \ \tilde \Phi_i$ in \cite{bfs,bc}.  Our discussion of the
LNN limit here extends the previous results in \cite{bfs,bc} (see also
\cite{gkgg,othersusy}). 

The $\beta$ function of this theory will be denoted $\beta_{\xi,s}$ and has the
expansion (\ref{beta}) with $\hat b_\ell$ and $\tilde b_\ell$ replaced by $\hat
b_{\ell,s}$ and $\tilde b_{\ell,s}$.  Here and below we use the subscript $s$,
standing for ``supersymmetric'', to avoid confusion with the corresponding
quantities discussed above in the nonsupersymmetric theory. Thus, 
\beq
  \hat b_{\ell,s} = \lim_{LNN} \frac{b_{\ell,s}}{N_c^\ell} \ , \quad
\tilde b_{\ell,s} = \lim_{LNN} \frac{\bar b_{\ell,s}}{N_c^\ell} \ .
\label{bellsrel}
\eeq
We denote the $n$-loop $\beta_{\xi,s}$ function as $\beta_{\xi,s,n\ell}$.  The
scheme-independent coefficients $b_{1,s}$ and $b_{2,s}$, were computed in
\cite{b1s} and \cite{b2s}, and $b_{3,s}$ was computed in \cite{b3s} in the
dimensional reduction ($\overline{DR}$) scheme \cite{dred}.

In the LNN limit, 
\beq
\hat b_{1,s} = 3-r \ , 
\label{b1s}
\eeq
\beq
\hat b_{2,s} = 2(3-2r) \ , 
\label{b2s}
\eeq
and, in the $\overline{DR}$ scheme, 
\beq
\hat b_{3,s} = 21-21r+4r^2 \ . 
\label{b3s}
\eeq

From these values of the coefficients of $\beta_{\xi,s}$, it follows that
\beq
r_{b1z,s} = 3
\label{rb1zs}
\eeq
and
\beq
r_{b2z,s} = \frac{3}{2} \ . 
\label{fb2zs}
\eeq
Asymptotic freedom requires $r < r_{b1z}$, and, in this range, the two-loop 
$\beta$ function has an IR zero for $r > r_{b2z}$, so the interval $I_{r,s}$
is
\beq
I_{r,s}: \ \frac{3}{2} < r < 3 \ . 
\label{irs}
\eeq
This IR zero occurs at the value 
\beqs
\xi_{IR,2\ell,s} & = & -\frac{\tilde b_{1,s}}{\tilde b_{2,s}} = 
-\frac{4\pi \hat b_{1,s}}{\hat b_{2,s}} \cr\cr
                 & = & \frac{2\pi(3-r)}{2r-3} \ . 
\label{xiir_2loop_susy}
\eeqs

The coefficient $\hat b_{3,s}$ vanishes at two values, 
\beq
r_{b3z,s,(1,2)} = \frac{21 \pm \sqrt{105}}{8} \ . 
\label{rb3zsplusminus}
\eeq
Numerically, $r_{b3z,s,1}=1.344$ and $r_{b3z,s,2}=3.906$, so that 
\beq
r_{b3z,s,1} < r_{b2z} \ , \quad r_{b3z,s,2} > r_{b1z} \ .
\label{rb3zrels}
\eeq
Since $b_{3,s} < 0$ for $r_{b3z,s,1} < r < r_{b3z,s,2}$, it follows that, 
in the $\overline{DR}$ scheme, 
\beq
\hat b_{3,s} < 0 \quad \forall \quad r \in I_r \ . 
\label{b3snegative}
\eeq
We list numerical values of the $\tilde b_{\ell,s}$ in Table 
\ref{btilde_nloop_values_susy}. 

\begin{figure}
  \begin{center}
    \includegraphics[height=14cm]{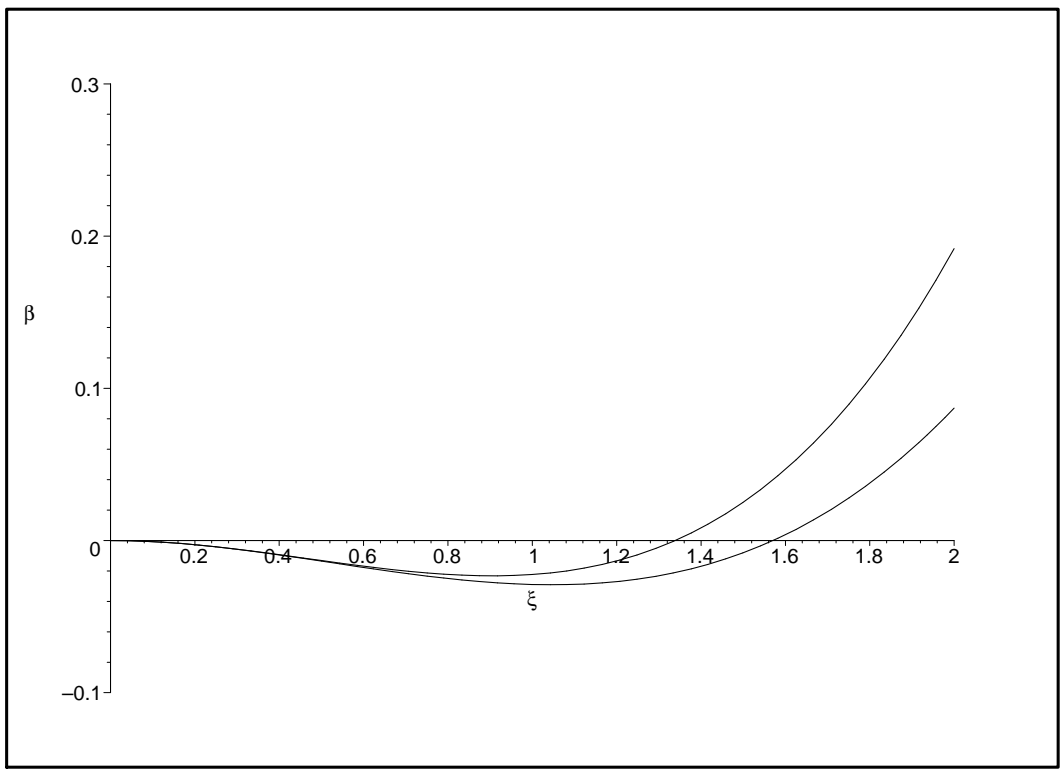}
  \end{center}
\caption{\footnotesize{Plot of the $n$-loop beta function,
 $\beta_{\xi,n\ell,s}$, as a function of $\xi$, for the illustrative value
 $r=2.5$ in the supersymmetric theory. From bottom to top, the curves represent
 $\beta_{\xi,2\ell,s}$ and $\beta_{\xi,3\ell,s}$, respectively.}}
\label{betaxi_susy_r2p5}
\end{figure}

The three-loop $\beta$ function formally vanishes at two points away from the
origin, at the zeros of $\tilde b_{1,s} + \tilde b_{2,s}\xi + \tilde
b_{3,s}\xi^2$, namely 
\beq
\xi = \frac{1}{2 \tilde b_{3,s}}\Big [ -\tilde b_{2,s} \pm 
\sqrt{\tilde b_{2,s}^2 -4\tilde b_{1,s} \tilde b_{3,s} } \ \Big ] \ . 
\label{xis3loop}
\eeq
By the same type of argument that was given above for the nonsupersymmetric
theory, one may argue that the inequality (\ref{b3snegative}) applies more
generally than just in the $\overline{DR}$ scheme.  In the present context,
this argument is that if the theory has an IR zero in the (scheme-independent)
two-loop $\beta$ function, it is reasonable to require that a scheme should
preserve the existence of this IR zero at higher-loop level, this requires that
$\tilde b_3 < 0 \ \forall r \in I_{r,s}$. This follows since $\tilde b_{2,s}
\to 0$ at the lower end of the interval $I_{r,s}$, so unless $\tilde b_{3,s} <
0$ in this interval, the quantity $\tilde b_{2,s}^2 -4\tilde b_{1,s} \tilde
b_{3,s}$ would become negative and the value of the zeros of $\beta$ at the
three-loop level in Eq. (\ref{xis3loop}) would be complex.  Given that $\tilde
b_3 < 0$ for $r \in I_{r,s}$, one may rewrite Eq. (\ref{xis3loop}) in terms of
positive quantities and pick out the relevant IR zero of $\beta_{\xi,s,3\ell}$
as
\beq
\xi_{IR,3\ell,s} = \frac{1}{2|\tilde b_{3,s}|} \Big [ -|\tilde b_{2,s}| + 
\sqrt{\tilde b_{2,s}^2+4 \tilde b_{1,s} |\tilde b_{3,s}| } \ \Big ] \ . 
\label{xiir_3loop_susyb}
\eeq
Explicitly, 
\beq
\xi_{IR,3\ell,s} = \frac{4\pi\Big [ -(2r-3) + \sqrt{C_s} \ \Big ]}{D_s}
\label{xiir_3loop_susy}
\eeq
where
\beq
C_s = -54+72r-29r^2+4r^3
\label{cs_susy}
\eeq
and
\beq
D_s = -21 + 21r -4r^2 \ . 
\label{ds_susy}
\eeq
Note that $D_s > 0$ for $r_{b3z,s,1} < r < r_{b3z,s,2}$ and hence for all $r
\in I_{r,s}$.  Furthermore, $C_s$ is positive-definite for $r \in I_{r,s}$
(the zeros of $C_s$ occur at $r=1.338$ and $r=2.956 \pm 1.163i$).  
By the same reasoning as was given before in \cite{bc} and above, we have 
the inequality
\beq
\xi_{IR,3\ell,s} < \xi_{IR,2\ell,s} \ . 
\label{xiir_32loop_inequality}
\eeq
We list numerical values of $\xi_{IR,n\ell,s}$ in Table 
\ref{xiir_nloop_values_susy} and show a plot of $\beta_{\xi,n\ell,s}$ for
$n=2$ and $n=3$ loops, evaluated at an illustrative value $r$ in $I_{r,s}$, 
namely, $r=2.5$ in Fig. \ref{betaxi_susy_r2p5}. 

% =====================================================================

\subsection{Other Structural Features of $\beta_{\xi,n\ell,s}$ }

We also briefly discuss some structural features of $\beta_{\xi,s}$. The
two-loop $\beta$ function, $\beta_{\xi,s,2\ell}$, reaches a minimum on the 
interval $I_{r,s}$ at the value 
\beq
\xi_{m,2\ell,s} = \frac{4\pi(3-r)}{2r-3} \ . 
\label{xi_critical_2loop_susy}
\eeq
The ratio of this position of the minimum in the beta function relative to the
position of the IR zero is the same as in the nonsupersymmetric theory, namely
\beq
\frac{\xi_{m,2\ell,s}}{\xi_{IR,2\ell,s}} = \frac{2}{3} \ . 
\label{xi_critical_23loop_ratio_susy}
\eeq

The value of $\beta_{\xi,2\ell,s}$ at the minimum is 
\beq
(\beta_{\xi,2\ell,s})_{min} = -\frac{8\pi(3-r)^3}{27(2r-3)^2} \ . 
\label{betaxi_2loop_min_susy}
\eeq

The derivative $d\beta_{\xi,IR,2\ell,s}$ evaluated at $\xi=\xi_{IR,2\ell,s}$,
is 
\beq
\beta'_{\xi,IR,2\ell,s} = \frac{(3-r)^2}{2r-3} \ . 
\label{dbetaxi_2loop_dxi_at_xiir_2loop_susy}
\eeq
Corresponding expressions can be given at the three-loop level, but we proceed
now to analyze a quantity of considerable interest, namely the anomalous
dimension of $\Phi \tilde \Phi$.

% ====================================================================

\subsection{Anomalous Dimension} 

We next consider the LNN limit of the anomalous dimension of the
(gauge-invariant) quadratic chiral superfield product, $\Phi \tilde \Phi$,
denoted $\gamma_s \equiv \gamma_{m,s}$.  This is given by Eq. (\ref{gamma})
with the replacements $\hat c_\ell \to \hat c_{\ell,s}$ and $\tilde c_\ell \to
\tilde c_{\ell,s}$, and similarly for the $n$-loop expression,
$\gamma_{m,s,n\ell}$.  From the known results for $c_{1,s}$ and, in the
$\overline{DR}$ scheme, $c_{2,s}$ and $c_{3,s}$, we find, in the LNN limit,
\beq
\hat c_{1,s} = 2 
\label{c1hat_susy}
\eeq
\beq
\hat c_{2,s} = 2(2-r) 
\label{c2hat_susy}
\eeq
and
\beq
\hat c_{3,s} = 10 - 6r[1+2\zeta(3)]-2r^2  \ . 
\label{c3hat_susy}
\eeq
Values of the corresponding $\tilde c_{\ell,s}$ are listed as a function of $r$
in Table \ref{ctilde_nloop_values_susy}. 

The two-loop anomalous dimension, evaluated at the two-loop IR zero of
$\beta_{\xi,s}$, is \cite{bfs} 
\beq
\gamma_{_{IR,2\ell,s}} = \frac{r(r-1)(3-r)}{2(2r-3)^2}  \ . 
\label{gamma_ir_2loop_susy}
\eeq
The quantity $\gamma_{_{IR,2\ell,s}}$ is a monotonically
decreasing function of $r$ in the interval $I_{r,s}$.  It exceeds the upper
bound of 1 from conformal symmetry if $r < 2$, and hence, as was noted in
\cite{bfs}, in the interval $3/2 < r < 2$, the perturbative two-loop
calculation that yields $\gamma_{_{IR,2\ell,s}}$ gives an unphysical result and
is unreliable.  One can apply this upper bound because one knows from exact
results \cite{seiberg} that for $3/2 < r < 3$, the theory flows in the infrared
to a conformal, non-Abelian Coulomb phase.

The three-loop anomalous dimension, evaluated at the three-loop IR zero of
$\beta_{\xi,s}$, is 
\beq
\gamma_{_{IR,3\ell,s}} = \frac{2( A_s + B_s \sqrt{C_s} \ )}{D_s^3} \ , 
\label{gamma_ir_3loop_susy}
\eeq
where $C_s$ and $D_s$ were defined above in Eqs. 
(\ref{cs_susy}) and (\ref{ds_susy}), and $A_s$ and $B_s$ are given in the
appendix.  
In contrast to $\gamma_{_{IR,2\ell,s}}$, $\gamma_{_{IR,3\ell,s}}$ is
not a monotonic function of $r$.  It reaches a maximum of approximately 0.1376
at $r=2.474$ and vanishes not just at $r=r_{b1z}=3$, but also at $r=2.1794$ (to
the indicated number of significant figures).  These features were evident for
the specific values of $N_c$ considered in \cite{bfs}; here we have shown how
this occurs after the LNN limit is taken. In Table
\ref{gamma_ir_nloop_values_susy} we list values of $\gamma_{_{IR,n\ell,s}}$ for
a range of $r$ values in $I_{r,s}$.  We note that for $1.5 < r < 2.0$,
$\gamma_{_{IR,2\ell,s}}$ exceeds the upper bound of 1 and hence is unphysical;
we do not include entries for these values of $r$.

% =======================================================================
\section{Discussion and Conclusions}
\label{conclusions} 

In summary, we have studied higher-loop corrections to the UV to IR evolution
of an asymptotically free vectorial SU($N_c$) gauge theory with $N_f$ fermions
in the fundamental representation, in the 't Hooft-Veneziano (LNN) limit $N_c
\to \infty$ and $N_f \to \infty$ with $r=N_f/N_c$ fixed and
$\xi(\mu)=\alpha(\mu) N_c$, a function independent of $N_c$ in this limit. We
have defined a beta function, $\beta_\xi$, that is finite in this LNN limit and
have analyzed its properties for the interval of $r$ in which
$\beta_{\xi,2\ell}$ has an IR zero.  We have given analytic and numerical
results for the LNN limiting quantities $\xi_{IR,n\ell}$, $\xi_{m,n\ell}$,
$(\beta_{\xi,n\ell})_{min}$, $\beta'_{\xi,IR,n\ell}$, and $\gamma_{IR,n\ell}$
as functions of $r$.  We have argued that a reasonable scheme should preserve
at higher loops the IR zero that is present at the (scheme-independent)
two-loop order in $\beta$, and that this implies that $\tilde b_3 < 0$ for $r
\in I_r$. In turn, this implies that $\xi_{IR,3\ell} < \xi_{IR,2\ell}$.
Calculating with the $\overline{MS}$ scheme, we find that in the part of the
interval $I_r$ where the perturbative calculations are reliable, the change in
the IR zero is smaller in magnitude going from three-loop to four-loop order,
as compared with the shift from two-loop to three-loop order.  This is in
agreement with one's expectation, that insofar as perturbative methods are
trustworthy, when a quantity is calculated to higher orders, the successive
changes should become smaller.  Further, we find that for the range of $r \in
I_r$ where the three-loop anomalous dimension is reliably calculable,
$\gamma_{IR,3\ell} < \gamma_{IR,2\ell}$.  These higher-loop calculations allow
one to extend the analysis of the IR zero of the $\beta$ function, and
corresponding evaluations of $\gamma_m$ to smaller values of $r$ and thus
stronger couplings than is possible with the two-loop result.  We have analyzed
the correction terms to the LNN-limit values of a number of quantities,
including $b_\ell/N_c^\ell$, $c_\ell/N_c^\ell$, $\alpha_{IR,n\ell}N_c$,
$\alpha_{m,n\ell}N_c$, $(\beta_{\alpha,n\ell})_{min}N_c$,
$d\beta_{\alpha,n\ell}/d\alpha {}|_{\alpha_{IR,n\ell}}$, and
$\gamma_{IR,n\ell}$, and have shown that these correction terms are suppressed
by $1/N_c^2$. This provides an understanding of the approximate universality
that is exhibited in calculations of these quantities for different values of
$N_c$ and $N_f$ with similar or identical values of $r$, even for moderate
values of $N_c$ and $N_f$.  A corresponding analysis was also given of a
vectorial gauge theory with ${\cal N}=1$ supersymmetry and $N_f$ chiral
superfields transforming according to the fundamental and conjugate fundamental
representation of SU($N_c$).  Thus, in addition to being of interest in its own
right, the LNN limit is useful in understanding common features of the UV to IR
evolution of various theories with different values of $N_c$ and $N_f$.

\bigskip
\bigskip

% ======================================================================

Acknowledgments: I would like to thank T. Ryttov for collaboration on the
earlier works \cite{bvh,bfs,sch}, and T. Appelquist and the theory group at
Yale University for warm hospitality during the sabbatical period when some of
this work was done.  This research was partially supported by the grant
NSF-PHY-09-69739.

\bigskip
\bigskip
\bigskip

% =======================================================================

\section{Appendix}

In this appendix we list some of the more lengthy expressions that are used in
the text.  The functions $F_{3\ell}$ and $G_{3\ell}$ that enter in Eq.
(\ref{betaxi_3loop_min}) for $(\beta_{\xi,3\ell})_{min}$ are
\begin{widetext}
\beqs
& & F_{3\ell} = -10985980784 + 17408705952r - 10177907376r^2
+ 2883132208r^3  - 468256107r^4 + 42131712r^5 -1605632r^6 \cr\cr
& & 
\label{a3ell}
\eeqs
and
\beq
G_{3\ell} = 41737992 - 48660252r + 18696078r^2 - 2733297r^3 + 139776r^4   \ . 
\label{b3ell}
\eeq

For the coefficient $c_4$ in Eq. (\ref{gamma}) (with fermions in the
fundamental representation, and in the $\overline{MS}$ scheme), using
\cite{gamma4}, we find
\beqs
c_4 & = & \hat c_4 + \frac{1}{N_c^2} \bigg [ -\frac{21947}{24} + 
\frac{126689}{324}r - \frac{1127}{162}r^2 + \frac{83}{81}r^3 + 
\Big ( -\frac{221}{9} + \frac{379}{3}r - \frac{16}{9}r^2 \Big ) \zeta(3) 
- 60r\zeta(5) \bigg ] \cr\cr
& + & \frac{1}{N_c^4} \bigg [ \frac{108359}{288} - \frac{9143}{108}r 
+ \frac{38}{27} r^2 + (-151+59r-20r^2)\zeta(3) + r(-66+12r)\zeta(4) 
+ (220-160r)\zeta(5) \bigg ] \cr\cr
& + & \frac{1}{N_c^6} \bigg [ -\frac{5783}{24} - \frac{37}{3}r + 
(89+111r)\zeta(3) + 60r\zeta(5) \bigg ] + 
\frac{1}{N_c^8}\Big [ -\frac{1261}{64} - 42\zeta(3) \Big ] \ . 
\label{c4_lnn}
\eeqs

The functions $A_s$ and $B_s$ that enter in the expression for
$\gamma_{_{IR,3\ell,s}}$ are 
\beq
A_s = 918-2916r+4146r^2-3322r^3+1532r^4-378r^5+40r^6
    + r \Big (2754-5508r+3942r^2-1212r^3+144r^4 \ \Big ) \zeta(3) 
\label{as_susy}
\eeq
and
\beq
B_s = 54-75r+17r^2+7r^3+5r^4-4r^5 + 
  r \Big (162-216r+102r^2-24r^3 \Big ) \zeta(3) \ . 
\label{bs_susy}
\eeq 
\end{widetext}

\newpage

% table I
\begin{table}
\caption{\footnotesize{Values of the $\tilde b_\ell$ coefficients for $1 \le
\ell \le 4$ as functions of $r$ for $0 \le r \le r_{b1z}$. Notation $a$e-n
means $a \times 10^{-n}$ here and in the other tables.}}
\begin{center}
\begin{tabular}{|c|c|c|c|c|c|} \hline\hline
$r$ & $\tilde b_1$ & $\tilde b_2$ & $\tilde b_3$ & $\tilde b_4$ 
\\ \hline
 0.0  & 0.2918     &   0.7177e-1   &   0.2666e-1   &   0.1265e-1    \\
 0.5  & 0.2653     &   0.5805e-1   &   0.1895e-1   &   0.8063e-2    \\
 1.0  & 0.2387     &   0.4433e-1   &   0.1176e-1   &   0.4429e-2    \\
 1.5  & 0.2122     &   0.3061e-1   &   0.5091e-2   &   0.1766e-2    \\
 2.0  & 0.1857     &   0.1689e-1   & $-0.1055$e-2  &   0.9062e-4    \\
 2.5  & 0.1592     &   0.3166e-2   & $-0.6677$e-2  & $-0.5814$e-3   \\
 3.0  & 0.1326     & $-0.1055$e-1  & $-0.1178$e-1  & $-0.2340$e-3   \\
 3.5  & 0.1061     & $-0.2427$e-1  & $-0.1635$e-1  &   0.1149e-2    \\
 4.0  & 0.7958e-1  & $-0.3800$e-1  & $-0.2041$e-1  &   0.3584e-2    \\
 4.5  & 0.5305e-1  & $-0.5172$e-1  & $-0.2394$e-1  &   0.7086e-2    \\
 5.0  & 0.2653e-1  & $-0.6544$e-1  & $-0.2695$e-1  &   0.1167e-1    \\
 5.5  & 0          & $-0.7916$e-1  & $-0.2944$e-1  &   0.1736e-1    \\
\hline\hline
\end{tabular}
\end{center}
\label{btilde_nloop_values}
\end{table}

% table II
\begin{table}
\caption{\footnotesize{Values of the $\tilde b_\ell$ with $1 \le \ell \le 4$ at
special values of $r$, including $r=0$ and at the lower and upper ends of the
interval $I_r$, $r=r_{b2z}=34/13$, and $r=r_{b1z}=11/2$.}}
\begin{center}
\begin{tabular}{|c|c|c|c|} \hline\hline
$\ell$ & $(\tilde b_\ell)_{r=0}$ & $(\tilde b_\ell)_{r=r_{b2z}}$ &
$(\tilde b_\ell)_{r=r_{b1z}}$ \\ \hline
1 & 0.2918     & 0.1530          &  0              \\
2 & 0.7177e-1  & 0               & $-0.7916$e-1    \\
3 & 0.2666e-1  & $-0.7900$e-2    & $-0.2944$e-2    \\
4 & 0.1265e-1  & $-0.5923$e-3  &  0.1736e-1        \\
\hline\hline
\end{tabular}
\end{center}
\label{btilde_nloop_specialvalues}
\end{table}

% table III 
\begin{table}
\caption{\footnotesize{Values of the IR zero $\xi_{IR,n\ell}$
of $\beta_{\xi,n\ell}$ function for $n=2, \ 3, \ 4$ and $r \in I_r$.}}
\begin{center}
\begin{tabular}{|c|c|c|c|} \hline\hline
$r$ & $\xi_{IR,2\ell}$ & $\xi_{IR,3\ell}$ & $\xi_{IR,4\ell}$
\\ \hline
 2.8   &  28.274     &  3.573      &  3.323       \\
 3.0   &  12.566     &  2.938      &  2.868       \\
 3.2   &  7.606      &  2.458      &  2.494       \\
 3.4   &  5.174      &  2.076      &  2.168       \\
 3.6   &  3.731      &  1.759      &  1.873       \\
 3.8   &  2.774      &  1.489      &  1.601       \\
 4.0   &  2.095      &  1.252      &  1.349       \\
 4.2   &  1.586      &  1.041      &  1.115       \\
 4.4   &  1.192      &  0.8490     &  0.9003      \\
 4.6   &  0.8767     &  0.6725     &  0.7038      \\
 4.8   &  0.6195     &  0.5083     &  0.5244      \\
 5.0   &  0.4054     &  0.3538     &  0.3603      \\
 5.2   &  0.2244     &  0.2074     &  0.2089      \\
 5.4   &  0.06943    &  0.06769    &  0.06775     \\
\hline\hline
\end{tabular}
\end{center}
\label{xiir_nloop_values}
\end{table}

% fig. 1
%
\begin{figure}
  \begin{center}
    \includegraphics[height=14cm]{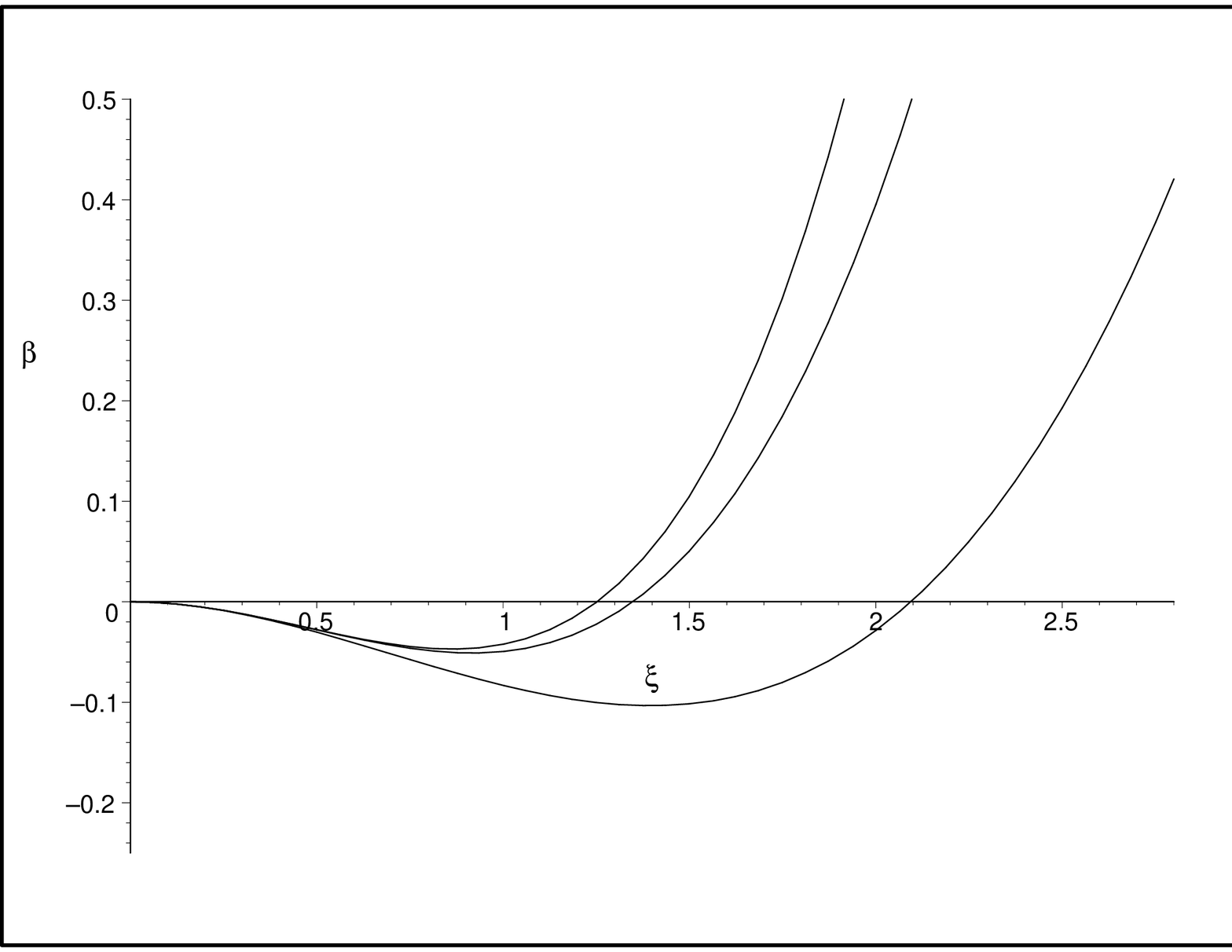}
  \end{center}
\caption{\footnotesize{Plot of the $n$-loop beta function,
 $\beta_{\xi,n\ell}$, as a function of $\xi$, for the illustrative value
 $r=4$. From bottom to top, the curves represent $\beta_{\xi,2\ell}$, 
$\beta_{\xi,4\ell}$, and $\beta_{\xi,3\ell}$, respectively.}}
\label{betaxi_r4}
\end{figure}
%

% table IV 
\begin{table}
\caption{\footnotesize{Values of the $\tilde c_\ell$ coefficients with $1 \le
\ell \le 4$ as functions of $r$ for $0 \le r \le r_{b1z}$. Here $\tilde
c_1=3/(4\pi)= 0.2387$, independent of $r$.}}
\begin{center}
\begin{tabular}{|c|c|c|c|c|c|} \hline\hline
$r$ & $\tilde c_2$ & $\tilde c_3$ & $\tilde c_4$ 
\\ \hline
 0.0  &  0.1071      &   0.5325e-1    &   0.2908e-1    \\
 0.5  &  0.1018      &   0.4396e-1    &   0.2098e-1    \\
 1.0  &  0.9657e-1   &   0.3434e-1    &   0.1324e-1    \\
 1.5  &  0.9129e-1   &   0.2440e-1    &   0.5897e-2    \\
 2.0  &  0.8602e-1   &   0.1413e-1    & $-0.1010$e-2   \\
 2.5  &  0.8074e-1   &   0.3538e-2    & $-0.7451$e-2   \\
 3.0  &  0.7546e-1   & $-0.7384$e-2   & $-0.1339$e-1   \\
 3.5  &  0.7019e-1   & $-0.1863$e-1   & $-0.1880$e-1   \\
 4.0  &  0.6491e-1   & $-0.3021$e-1   & $-0.2364$e-1   \\
 4.5  &  0.5963e-1   & $-0.4211$e-1   & $-0.2788$e-1   \\
 5.0  &  0.5435e-1   & $-0.5434$e-1   & $-0.3148$e-1   \\
 5.5  &  0.4908e-1   & $-0.6690$e-1   & $-0.3442$e-1   \\
\hline\hline
\end{tabular}
\end{center}
\label{ctilde_nloop_values}
\end{table}

% table V
\begin{table}
\caption{\footnotesize{Values of the $n$-loop anomalous dimension,
$\gamma_{n\ell}$, evaluated at the $n$-loop IR zero of $\beta_\xi$ and
denoted $\gamma_{_{IR,n\ell}}$, as in Eq. (\ref{gamma_xi_ir_nloop}), where
$n=2, \ 3, \ 4$, for $r \in I_r$.  In the entries marked u, the $n$-loop
perturbative value of $\gamma_{IR,n\ell}$ is larger than the upper bound of 2
and hence is unphysical (u).}}
\begin{center}
\begin{tabular}{|c|c|c|c|} \hline\hline
$r$ & $\gamma_{_{IR,2\ell}}$ & $\gamma_{_{IR,3\ell}}$ & $\gamma_{_{IR,4\ell}}$
\\ \hline
 2.8   &  u        &  1.708      &  0.1902  \\
 3.0   &  u        &  1.165      &  0.2254  \\
 3.2   &  u        &  0.8540     &  0.2637  \\
 3.4   &  u        &  0.6563     &  0.2933  \\
 3.6   &  1.853    &  0.5201     &  0.3083  \\
 3.8   &  1.178    &  0.4197     &  0.3061  \\
 4.0   &  0.7847   &  0.3414     &  0.2877  \\
 4.2   &  0.5366   &  0.2771     &  0.2664  \\
 4.4   &  0.3707   &  0.2221     &  0.2173  \\
 4.6   &  0.2543   &  0.1735     &  0.1745  \\
 4.8   &  0.1696   &  0.1294     &  0.1313  \\
 5.0   &  0.1057   &  0.08886    &  0.08999 \\
 5.2   &  0.05620  &  0.05123    &  0.05156 \\
 5.4   &  0.01682  &  0.01637    &  0.01638 \\
\hline\hline
\end{tabular}
\end{center}
\label{gamma_ir_nloop_values}
\end{table}

% ===================== susy results =============================

% table VI 
\begin{table}
\caption{\footnotesize{Values of the $\tilde b_{\ell,s}$ coefficients for $1
\le \ell \le 3$ as functions of $r$ for $0 \le r \le r_{b1z,s}$ in the
supersymmetric theory.}}
\begin{center}
\begin{tabular}{|c|c|c|c|c|} \hline\hline
$r$ & $\tilde b_{1,s} $ & $\tilde b_{2,s}$ & $\tilde b_{3,s}$ 
\\ \hline
 0.0  & 0.2387     &   0.3800e-1   &   0.1058e-1   \\
 0.5  & 0.1989     &   0.2533e-1   &   0.5795e-2   \\
 1.0  & 0.1592     &   0.1267e-1   &   0.20166e-2  \\
 1.5  & 0.1194     &   0           & $-0.7559$e-3  \\
 2.0  & 0.7958e-1  & $-0.1267$e-1  & $-0.2520$e-2  \\
 2.5  & 0.3979e-1  & $-0.2533$e-1  & $-0.3276$e-2  \\
 3.0  & 0          & $-0.3800$e-1  & $-0.3024$e-2  \\
\hline\hline
\end{tabular}
\end{center}
\label{btilde_nloop_values_susy}
\end{table}

% table VII 
\begin{table}
\caption{\footnotesize{Values of the IR zero $\xi_{IR,n\ell,s}$
of $\beta_{\xi,n\ell,s}$ function for $n=2$ and $n=3$ loops and
$r \in I_{r,s}$.}}
\begin{center}
\begin{tabular}{|c|c|c|c|} \hline\hline
$r$ & $\xi_{IR,2\ell,s}$ & $\xi_{IR,3\ell,s}$ 
\\ \hline
 1.8   &  12.566     &  5.331    \\
 1.9   &   8.639     &  4.381    \\
 2.0   &   6.283     &  3.643    \\
 2.1   &   4.712     &  3.040    \\
 2.2   &   3.590     &  2.529    \\
 2.3   &   2.749     &  2.085    \\
 2.4   &   2.094     &  1.692    \\
 2.5   &   1.571     &  1.339    \\
 2.6   &   1.142     &  1.019    \\
 2.7   &   0.7854    &  0.7279   \\
 2.8   &   0.4833    &  0.4623   \\
 2.9   &   0.2244    &  0.2201   \\
\hline\hline
\end{tabular}
\end{center}
\label{xiir_nloop_values_susy}
\end{table}

% fig. 2
%
%

% table VIII 
\begin{table}
\caption{\footnotesize{Values of the $\tilde c_{\ell,s}$ coefficients 
for $\ell=2$ and $\ell=3$, as functions of $r$ for $0 \le r \le r_{b1z,s}$ 
in the supersymmetric theory. Here $\tilde c_{1,s}=1/(2\pi)=0.1592$,
independent of $r$.}}
\begin{center}
\begin{tabular}{|c|c|c|} \hline\hline
$r$   & $\tilde c_{2,s}$ & $\tilde c_{3,s}$ \\ \hline
 0.0  &   0.2533e-1      &   0.5039e-2   \\
 0.5  &   0.1900e-1      & $-0.3590$e-3  \\
 1.0  &   0.1267e-1      & $-0.6261$e-2  \\
 1.5  &   0.6333e-2      & $-0.1267$e-1  \\
 2.0  &   0              & $-0.1958$e-1  \\
 2.5  & $-0.6333$e-2     & $-0.2699$e-1  \\
 3.0  & $-0.1267$e-1     & $-0.3491$e-1  \\
\hline\hline
\end{tabular}
\end{center}
\label{ctilde_nloop_values_susy}
\end{table}

% table IX
\begin{table}
\caption{\footnotesize{Values of the $n$-loop anomalous dimension,
$\gamma_{n\ell,s}$, evaluated at the $n$-loop IR zero of 
$\beta_{\xi,s}$ and denoted $\gamma_{_{IR,n\ell,s}}$, where
$n=2$ and $n=3$, for $r \in I_{r,s}$.  For $1.5 < r < 2.0$, 
$\gamma_{_{IR,2\ell,s}} > 1$ and hence is unphysical (u).}}
\begin{center}
\begin{tabular}{|c|c|c|} \hline\hline
$r$ & $\gamma_{IR,2\ell,s}$ & $\gamma_{IR,3\ell,s}$ 
\\ \hline
 1.8   &  u          & $-1.617$      \\
 1.9   &  u          & $-0.8053$     \\
 2.0   &  1          & $-0.3667$     \\
 2.1   &  0.7219     & $-1183$       \\
 2.2   &  0.5388     &  0.2267e-1    \\
 2.3   &  0.4088     &  0.9809e-1    \\
 2.4   &  0.3111     &  0.1314       \\
 2.6   &  0.2344     &  0.1370       \\
 2.7   &  0.1719     &  0.9955e-1    \\
 2.8   &  0.7456e-1  &  0.6829e-1    \\
 2.9   &  0.3514e-1  &  0.3412e-1    \\
\hline\hline
\end{tabular}
\end{center}
\label{gamma_ir_nloop_values_susy}
\end{table}


\begin{thebibliography}{99}

% 1
\bibitem{beta}
%
C. G. Callan, Phys. Rev. D {\bf 2}, 1541 (1970); K. Symanzik, Commun. Math.
Phys. {\bf 18}, 227 (1970).  See also M. Gell-Mann and F. Low, Phys. Rev. {\bf
95}, 1300 (1954); N. N. Bogolubov and D. V. Shirkov, Doklad. Akad. Nauk SSSR
{\bf 103}, 391 (1955); K. Wilson, Phys. Rev. D {\bf 3}, 1818 (1971).

% 2
\bibitem{fm} 
%
Our restriction to massless fermions is only for technical
convenience.  It is easy to include fermion mass terms, which are
gauge-invariant in a vectorial gauge theory.  However, if a given fermion has a
mass $m$, it is integrated out of the effective field theory applicable at
scales $\mu < m$ and does not affect the evolution in this region.

% 3
\bibitem{b1}
D. J. Gross and F. Wilczek, Phys. Rev. Lett. {\bf 30}, 1343 (1973);
H. D. Politzer, Phys. Rev. Lett. {\bf 30}, 1346 (1973); G. 't Hooft,
unpublished.

% 4
\bibitem{b2}
W. E. Caswell, Phys. Rev. Lett. {\bf 33}, 244 (1974);
D. R. T. Jones, Nucl. Phys. B {\bf 75}, 531 (1974).

% 5
\bibitem{bz}
T. Banks and A. Zaks, Nucl. Phys. B {\bf 196}, 189 (1982).

% 6
\bibitem{b3}
O. V. Tarasov, A. A. Vladimirov, and A. Yu. Zharkov, Phys. Lett. B {\bf 93},
429 (1980); S. A. Larin and J. A. M. Vermaseren, Phys. Lett. B {\bf 303}, 334
(1993).

% 7
\bibitem{b4}
T. van Ritbergen, J. A. M. Vermaseren, and S. A. Larin, Phys. Lett. B {\bf
400}, 379 (1997).

% 8
\bibitem{gkgg}
E. Gardi and M. Karliner, Nucl. Phys. B {\bf 529}, 383 (1998);
E. Gardi and G. Grunberg, JHEP 03, 024 (1999).

% 9
\bibitem{bvh}
T. A. Ryttov and R. Shrock, Phys. Rev. D {\bf 83}, 056011 (2011),
arXiv:1011.4542.

% 10
\bibitem{ps}
C. Pica and R. Sannino, Phys. Rev. D {\bf 83}, 035013 (2011),
arXiv:1011.5917.

% 11
\bibitem{bfs}
T. A. Ryttov, R. Shrock, Phys. Rev. D {\bf 85}, 076009 (2012), 
arXiv:1202.1297.  

% 12
\bibitem{bc}
R. Shrock, arXiv:1301.3209. 

% 13
\bibitem{thooftlargeN}
G. 't Hooft, Nucl. Phys. B {\bf 72}, 461 (1974), Nucl. Phys. B {\bf 75}, 
461 (1974).

% 14
\bibitem{veneziano}
G. Veneziano, Nucl. Phys B {\bf 117}, 519 (1976). 

% 15
\bibitem{lnc}
C.-K. Chow and T.-M. Yan, Phys. Rev. D {\bf 53}, 5105 (1996); 
R. Shrock, Phys. Rev. D {\bf 53}, 6465 (1996); R. Shrock, Phys. Rev.
D {\bf 76}, 055010 (2007).

% 16
\bibitem{stanley}
H. E. Stanley, Phys. Rev. {\bf 176}, 718 (1968); Phys. Rev. {\bf 179}, 570 
(1969). 

% 17
\bibitem{earlylargeN}
%
Some early applications of the large-$N$ limit to quantum field theories
include 
H. J. Schnitzer, Phys. Rev. D {\bf 10}, 1800 (1974); 
S. R. Coleman, R. Jackiw, and H. D. Politzer, Phys. Rev. D {\bf 10}, 2491 
(1974); 
D. J. Gross and A. Neveu, Phys. Rev. D {\bf 10}, 3235 (1974); 
C. G. Callan, N. Coote, and D. G. Gross, Phys. Rev. D {\bf 13}, 1649 (1976);
E. Br\'ezin and J. Zinn-Justin, Phys. Rev. B {\bf 14}, 3110 (1976);
W. A. Bardeen, B. W. Lee, and R. E. Shrock, Phys. Rev. D {\bf 14}, 985 (1976);
M. B. Einhorn, S. Nussinov, and E. Rabinovici, Phys. Rev. D {\bf 15}, 
2282 (1977); J. Koplik, A. Neveu, S. Nussinov, Nucl. Phys. B {\bf 123}, 109 
(1977); R. C. Brower, J. Ellis, M. G. Schmidt, and J. H. Weis, 
Nucl. Phys. {\bf 128}, 131 (1977); 
B. De Wit and G. 't Hooft, Phys. Lett. B {\bf 69}, 61 (1977); 
T. T. Wu, Phys. Lett. B {\bf 71}, 142 (1977); 
E. Br\'ezin, C. Itzykson, G. Parisi, and J.-B. Zuber, Commun. Math. 
Phys. {\bf 59}, 35 (1978); E. Witten, Nucl. Phys. B {\bf 160}, 57 (1979); 
E. Corrigan and P. Ramond, Phys. Lett. B {\bf 87}, 73 (1979);
Yu. Makeenko and A. A. Migdal, Phys. Lett. B {\bf 88}, 135 (1979);
S. R. Coleman and E. Witten, Phys. Rev. Lett. {\bf 45}, 100 (1980);
H. Neuberger, Phys. Lett. B {\bf 94}, 199 (1980). 

% 18 
\bibitem{largeNreview}
%
For a recent review with references to the extensive literature on large-$N$
methods, see B. Lucini and M. Panero, Phys. Repts., in press (2013).

% 19
\bibitem{casimir}
%
The Casimir invariants $C_R$ and $T_R$ are defined as $\sum_a \sum_j {\cal
D}_R(T_a)_{ij} {\cal D}_R(T_a)_{jk} = C_R \delta_{ik}$ and $\sum_{i,j} {\cal
D}_R(T_a)_{ij} {\cal D}_R(T_b)_{ji} = T_R \delta_{ab}$, where $R$ is the
representation and $T_a$ are the generators of $G$, so that for SU($N_c$),
$C_A=N_c$ for the adjoint ($A$) and $T_{fund}=1/2$ for the
fundamental representation, etc. $C_f$ denotes $C_R$ for the fermion 
representation. 

% 20
\bibitem{ap}
T. Appelquist, J. Terning, and L. C. R. Wijewardhana, Phys. Rev. Lett. 
{\bf 77}, 1214 (1996). 

% 21
\bibitem{conf}
%
See \cite{bvh,bc} for further references on this. For recent reviews of lattice
and continuum studies of this chiral transition for various $N_c$, fermion
representations, and $N_f$, see, e.g., the talks at the conferences {\it
Lattice Meets Experiment 2012: Beyond the Standard Model}, Univ. of Colorado,
Oct., 2012, URL http://www-hep.colorado.edu/~schaich/lat-exp-2012 and
Proceedings of {\it Strongly Coupled Gauge Theories in the LHC Perspective},
Dec., 2012, Univ. of Nagoya, ed. K. Yamawaki, to appear.

% 22
\bibitem{ms}
G. 't Hooft, Nucl. Phys. B {\bf 61}, 455 (1973).

% 23
\bibitem{msbar}
W. A. Bardeen, A. J. Buras, D. W. Duke, and T. Muta, Phys. Rev. D {\bf 18},
3998 (1978).

% 24
\bibitem{bethke}
S. Bethke, Eur. Phys. J. C {\bf 64}, 689 (2009).

% 25
\bibitem{sch}
T. A. Ryttov and R. Shrock,
Phys. Rev. D {\bf 86}, 065032 (2012), arXiv:1206.2366;
Phys. Rev. D {\bf 86}, 085005 (2012), arXiv:1206.6895.

% 26
\bibitem{gamma4}
J. A. M. Vermaseren, S. A. Larin, and T. van Ritbergen, Phys. Lett. B {\bf
  405}, 327 (1997).

% 27
\bibitem{cftbound}
S. Ferrara, R. Gatto, A. F. Grillo, Phys. Rev. D {\bf 9}, 3564 (1974); 
G. Mack, Commun. Math. Phys. {\bf 55}, 1 (1977); 
B. Grinstein, K. Intriligator, and I. Rothstein, Phys. Lett. B {\bf 662}, 
367 (2008). 

% 28
\bibitem{sanrev}
A recent review is F. Sannino, Acta Phys. Polon. B {\bf 40}, 3533 (2009).

% 29
\bibitem{nsvzbeta}
V. A. Novikov, M. A. Shifman, A. I. Vainshtein, and V. I. Zakharov, Nucl.
Phys. B {\bf 229}, 381 (1983); Nucl. Phys. B {\bf 277}, 426 (1986).

% 30
\bibitem{seiberg}
N. Seiberg, Phys. Rev. D {\bf 49}, 6857 (1994); Nucl. Phys. B {\bf 435},
129 (1995); K. A. Intriligator and N. Seiberg, Nucl. Phys. B {\bf 444}, 125
(1995).

% 31
\bibitem{othersusy}
R. Oehme, Phys. Lett. B {\bf 399}, 67 (1997);
M. T. Frandsen, T. Pickup, and M. Teper, Phys. Lett. B {\bf 695}, 231 (2011).

% 32
\bibitem{b1s}
D. R. T. Jones, Nucl. Phys. B {\bf 87}, 127 (1975).

% 33
\bibitem{b2s}
%
M. Machacek and M. Vaughn, Nucl. Phys. B {\bf 222}, 83 (1983);
A. J. Parkes and P. C. West, Phys. Lett. B {\bf 138}, 99 (1984);
Nucl. Phys. B {\bf 256}, 340 (1985);
D. R. T. Jones and L. Mezincescu, Phys. Lett. B {\bf 136}, 242 (1984);
Phys. Lett. B {\bf 138}, 293 (1984).

% 34
\bibitem{b3s}
%
R. V. Harlander, D. R. T. Jones, P. Kant, L. Mihaila, and M. Steinhauser,
JHEP 0612, 024 (2006);
R. Harlander, L. Mihaila, and M. Steinhauser, Eur. Phys. J.
C {\bf 63}, 383 (2009).

% 35 
\bibitem{dred}
%
W. Siegel, Phys. Lett. B {\bf 84}, 193 (1979); Phys. Lett. B {\bf 94}, 37
(1980); a recent discussion is W. St\"ockinger, JHEP 0503, 076 (2005).


\end{thebibliography}
\end{document}